\documentclass{IEEEtran}
\usepackage{cite}
\usepackage{amsmath,amssymb,amsfonts,amsthm}
\usepackage[hyphens]{url}
\usepackage{algorithm}
\usepackage{algorithmicx}
\usepackage{graphicx}
\usepackage{float}
\usepackage{textcomp}
\usepackage{xcolor}
\usepackage{url}
\usepackage[noend]{algpseudocode}
\usepackage{amsfonts}
\usepackage{url}
\usepackage{subfigure}
\usepackage{verbatim}

\theoremstyle{definition}

\theoremstyle{remark}

\title{A DLT enabled smart mask system to enable social compliance}
\author{ Lianna Zhao, Pietro Ferraro and Robert Shorten}

\begin{document}
\maketitle

\begin{abstract}
As Covid-19 remains a cause of concern, especially due to its mutations, wearing masks correctly and efficiently remains a priority in order to limit the spread of the disease.  In this paper we present a wearable smart-mask prototype using concepts from Internet of Things, Control Theory and Distributed Ledger Technologies. Its purpose is to encourage people to comply with social distancing norms, through the use of incentives.
The smart mask is designed to monitor Carbon Dioxide and Total Volatile Organic Compounds concentrations. The detected data is appended to a DAG-based DLT, named the IOTA Tangle. The IOTA Tangle ensures that the data is secure and immutable and acts as a communication backbone for the incentive mechanism. A hardware-in-the-loop simulation, based on indoor positioning, is developed to validate the effectiveness of the designed prototype.

\end{abstract} 

\section{Introduction}
Due to the emergence of Covid-19 and its mutations, it is reasonable to expect that masks and social distancing norms will still play a significant role in many societies across the globe. According to European Centre for Disease Prevention and Control, from 31 December 2019 to 17 March 2022, 458 179 120 cases of Covid-19 (in accordance with the applied case definitions and testing strategies in the affected countries) have been reported, including 6 058 022 deaths. If not properly controlled, the virus might once again spread across the population, leading to high mortality rates and hospitalizations, 
even with possible sequelae\footnote{https://www.bmj.com/content/376/bmj-2021-068414}. Even as the amount of infected people abates in some areas, the need to wear face-masks remains in many aspects of daily life, for example, in passenger planes, buses and trains. In such situations, enforcement of mask wearing is the responsibility of {\em observers}, such as flight attendants, rather than the mask wearer. Often, this leads to situations where either compliance is not enforced, or where an unreasonable burden is placed on these {\em observers}. In this paper we explore ways to encourage people to wear masks properly, especially in confined and crowded spaces, such as supermarkets, tubes and airplanes. Importantly, we wish to design mechanisms where compliance with mask wearing remains with the mask wearer - rather than with {\em observers}. To do this we build on our previous work done in \cite{ferraro2021personalised}. Here the authors discuss a general framework, based on control theory, with the aim of regulating compliance to social contracts\footnote{In this paper, we define social contracts as a number of rules or policies, designed to guide people's interaction with other people and social infrastructures} in the sharing economy domain. \newline

\begin{figure}[H]
\centering
\includegraphics[width=0.95\columnwidth]{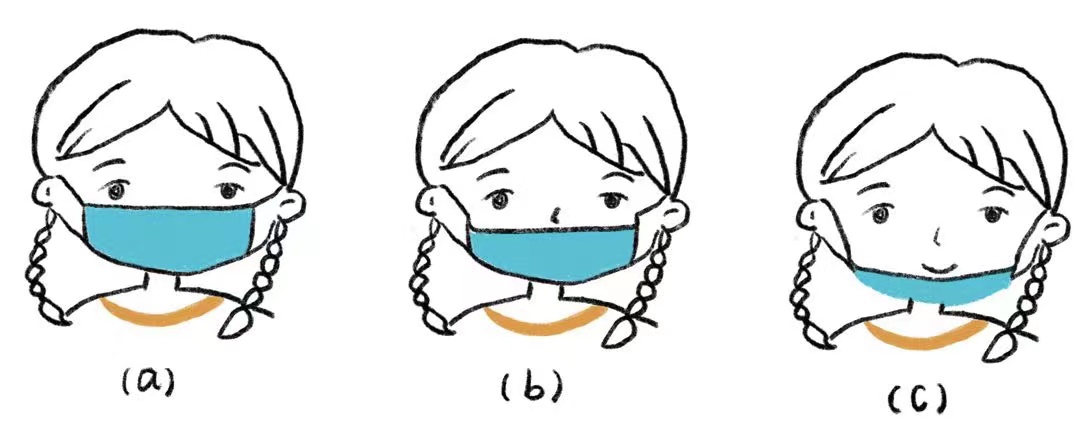}
\caption{The criterion to say people wear masks correctly or not.}
\label{fig: Maskwearing}
\end{figure}

The objective of this paper is to present a Proof-of-Concept (PoC) of a smart mask prototype that can detect people's mask-wearing status, and then incentivise people's behaviour with a token-bond mechanism to wear masks efficiently in confined and crowded spaces. By wearing masks correctly, we mean that people use mask to cover both their mouth and nose at the same time as shown in Figure \ref{fig: Maskwearing} (a), while Figure \ref{fig: Maskwearing} (b) and (c) are illustrations of not proper mask wearing.
While there are some algorithms that address  the compliance issue, such as game theoretic framework ( trying to design an equilibrium which encourages people's good behaviour), there are still some limitations within these works, for example, they are centralised; they are vulnerable to resource-rich attackers; they are often not anonymous because of unencrypted user address; fairness is not preserved as they are not tailored according to the individual situation.\newline
 
Our work also builds on the idea of a personalised dynamic pricing strategy. For completeness, we note that there are many paper on this topic; for example - see \cite{bui2012dynamic, phan2016model, kotb2016iparker}. Specifically, our work is based on \cite{ferraro2021personalised}, which differs from the works mentioned above along various dimensions. In \cite{ferraro2021personalised}, the authors propose a personalised feedback control based on distributed ledger technology (DLT) system to enforce people's compliance and furthermore provide a theoretical analysis on the designed system's convergence.  Here, DLT structure provides a more secure and privacy-preserving structure than its centralised alternatives, to create a personalised economic commitment algorithm and then to  enforce compliance. For the convenience of further discussing, we refer to the designed algorithm in \cite{ferraro2021personalised} as Personalised Feedback Control Algorithm (PFCA).
 The idea of incentives is also being explored in the circular economy and this concept has already been explored in various applications (although, not in a theoretical manner), such as Kupcrush. Kupcrush uses DLTs to make cups into their own economic agents through the use of Digital Twins. Each cup is associated with a digital identity and a wallet and their aim is to incentivise the consumer who is using the cup to recycle it correctly by 'rewarding every actor in the circular economy with a micro-reward as the cup moves through the recycling chain until it is ultimately recycled into a new cup'\footnote{https://challenges.dk/en/idea/kupkrush-single-use-paper-cup-circular-economy-recycling}.
 In this paper, we are proposing a variation on the same idea. We present a smart mask, in which we make use of a DLT structure to design a compliance strategy and enforce social contracts. Our proposal is to use digital tokens as a bond to encourage people's compliance: if people remain in compliance, they will get tokens back to their account
\cite{moschella2021decentralized, ferraro2018distributed, ferraro2019stability}, otherwise they will lose it.\newline


Accordingly, the main contributions of this paper are:\newline

\begin{itemize}
\item A smart mask with sensors is designed to detect people's mask wearing status. 
The recorded data is used to encourage compliance through a bond-deposit scheme implemented on the IOTA Tangle.\newline
\item A prototype based on a Raspberry Pi 3B \cite{kumar2017air, naik2019smart} hardware platform is presented.\newline
\item A hardware-in-the-loop simulation, based on indoor positioning, with ultra-wide band (UWB) is designed to validate the effectiveness of the proposed algorithm and the mask design.\newline
\end{itemize}

 The remainder of this paper is organised as follows: In Section II,  we give a brief description of DLT and provide a brief summary of the personalised feedback control Algorithm (PFCA) \cite{ferraro2021personalised}.
 In Section III, the mask design is introduced and the procedure of data analysis is described briefly. In Section IV we
illustrate the efficacy of proposed approach through hardware-in-the-loop simulation based on indoor positioning. Finally, in section V, we summarise the presented results and discuss future lines of research. \newline

\section{Background}

\subsection{IOTA Tangle}

In essence, DLT refers to digital ledgers shared across multiple nodes in a peer-to-peer network~\cite{kuhn2019rethinking}. It has recently gained popularity in both industry and academic communities; for example, in smart cities~\cite{ferraro2018distributed}, supply chain and health-care. DLTs hold great potential in these sectors because of their desirable properties, such as decentralization, immutability, consistency, and transparency. Although blockchain provides a decentralized architecture to overcome the shortcomings of the centralized architecture, several limitations in blockchain hinder its widespread application. For example, the inherent sequential structure in blockchain to add new transactions gives rise to its low scalability and the heavy cryptography Proof-of-Work (PoW) consensus leads to expensive computation power expenses and high transaction fees.\newline

As an alternative, the IOTA DLT, whose structure is Directed Acyclic Graph, is proposed in \cite{benvcic2018distributed}\cite{popov2018tangle}. Instead of using a chain, every new incoming transaction can freely reference existing transactions in a graph structure, without being subject to the restrictions as imposed by the blockchain. This means many transactions are verified in a parallel fashion. Every new transaction must approve two previous transactions. In the IOTA Tangle, there is no Proof-of-work and no transaction fees required. This later feature makes the IOTA DLT attractive for appliance applications.\newline


\subsection{PFCA}
\begin{figure}
\includegraphics[width=1\columnwidth]{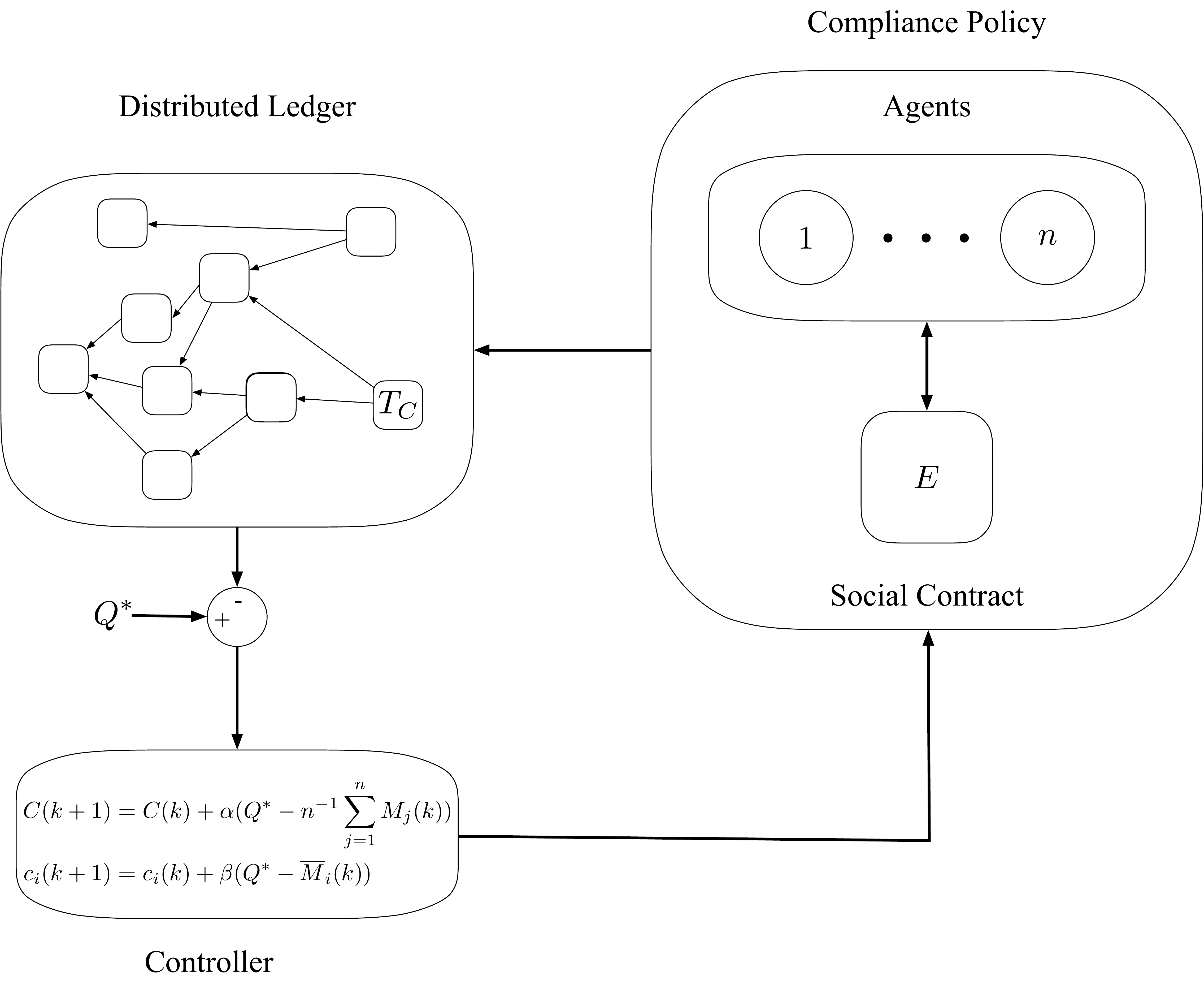}
\caption{The three components of the proposed compliance architecture are: the Distributed Ledger that acts as the communication backbone of the infrastructure, the compliance policy and the feedback controller \cite{ferraro2021personalised}.}
\label{fig: architecture}
\end{figure}

Our work builds on \cite{ferraro2021personalised}. 
The main idea in \cite{ferraro2021personalised} is to use digital tokens
to encourage people to comply with a social contract. By social contract, we mean a set of guidelines that must be followed to ensure the proper utilization of a resource or object. For example, the agreement of wearing a mask correctly, in the context of Covid-19, is a social contract.  The basic idea is that agents deposit tokens as a bond when they put on a mask, in areas where it is expected for them to wear one (e.g., an airplane). These tokens are then returned in full
if the person does not remove the mask. A pricing algorithm is used to determine the number of tokens that are deposited (based on the level of previous compliance). The architecture for realising this system in \cite{ferraro2021personalised} is depicted in Figure \ref{fig: architecture}. 
The algorithm in \cite{ferraro2021personalised} is  organized  around  three functional components: the distributed ledger is used as a communication layer (i.e., to record the deposit and the withdrawal of the tokens); the physical layer represents the agents interaction in engaging with the social contract; the controller is used to adjust the amount of tokens deposited and achieve expected compliance level. By adopting smart contracts, the whole
process including deposit and return of the tokens can be  automatically operated.
All operations are recorded on a DLT which is immutable and data are shared anonymously among agents (since, each agent's identity is represented by an encrypted address). There are four policies that could be adopted in this context: 
\begin{itemize}
\item {\bf Fixed penalty policy: } Before participating in the social scheme each agent deposits a certain amount of tokens, the amount being set by the controller. When the action is completed or when the agent exits the scheme, all tokens are returned in the event that they complied with rule $E$; otherwise no tokens are returned to the agent. In the latter case the pricing algorithm continues to adjust the price based on both the agents' level of compliance and that of the network.\newline

\item{\bf Adaptive penalty policies:} Initially each agent deposits a certain amount of tokens, the amount being set by the controller. The contract is reissued at every time-step. 
At each time step, compliant agents retrieve their tokens, and stake new ones to continue the activity. Non-compliant agents lose all their tokens every time they do not comply. At all time steps, the pricing algorithm continues to adjust the price based on both the agents' level of compliance and that of the network.\newline

\item{\bf Adaptive penalty policies with return:} Initially each agent deposits a certain amount of tokens, the amount being set by the controller. The contract is reissued at every time-step. 
At each time step, compliant agents retrieve their tokens, and stake new ones to continue the activity. Non-compliant agents lose all their tokens every time they do not comply. If an agent that previously lost a token starts complying again, they will retrieve a portion of the lost tokens. At all time steps, the pricing algorithm continues to adjust the price based on both the agents' level of compliance and that of the network.\newline

\item {\bf Event driven policies: } Initially each agent deposits a certain amount of tokens. Whenever the agent fails to comply with rule $E$ the tokens are lost; in order to keep participating in the scheme the agent needs to deposit more tokens. In this version of the scheme the amount of tokens that are required varies as a bond changes value over time (again a smart contract can easily take care of the update process).\newline
\end{itemize}

\begin{figure*}
\centering
\includegraphics[width=16cm, height=7.5cm]{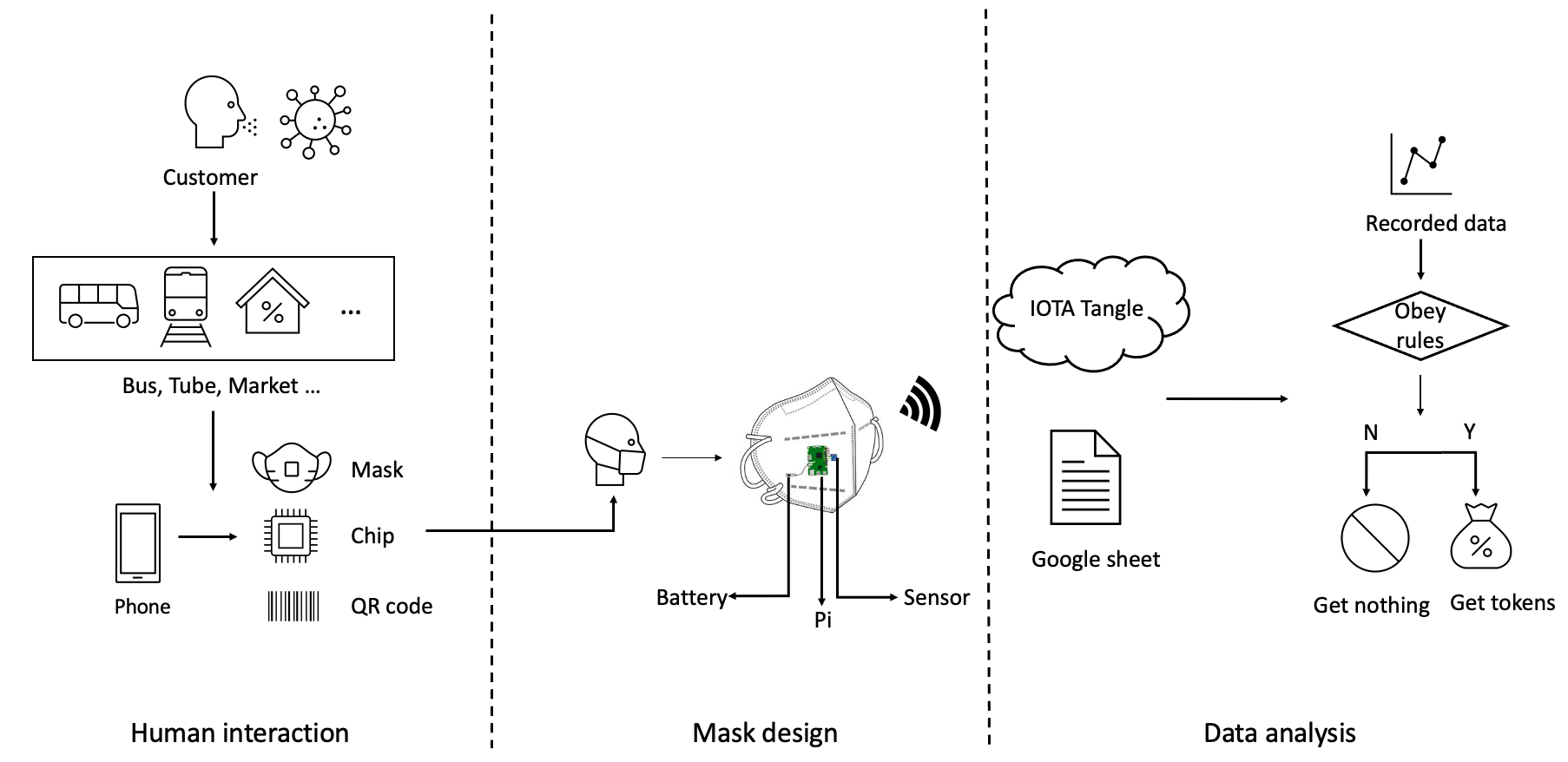}
\caption{The network structure.}
\label{fig: Network}
\end{figure*}

A feedback mechanism for designing a proper value of the bond can be constructed as follows. For each agent $i$, we define a  binary random variable $\{ M_i(k) \in \{0,1\} \}_{i=1}^n$ for discrete values $k$ as:
\begin{equation}
  P(\text{$i$ complies with rule $E$ at time $k$}) = P(M_i(k) = 1)
 \label{eq: prob}
\end{equation}

Moreover, we assume that the probability of these events is entirely dependant on a constant $q_i$, which represents the proclivity of each agent to comply with rules, and two control variables, $C(k)$, $c_i(k)$. The variable $C(k)+c_i(k)$ represents the value of the token bond staked by agent $i$ at time-step $k$. The combination $q_i + C(k)+c_i(k)$ determines the likelihood that agent $i$ will comply with the rule at time-step $k+1$. Then, (\ref{eq: prob}) can be expressed as

\begin{equation}
    \P(M_i(k+1) = 1) = p\left(q_i + C(k)+c_i(k)\right)
\end{equation}

with $p: \mathbb{R} \longrightarrow [0,1]$ being a monotone increasing function (which is used to bind the probability between 0 and 1). $C(k)$ and $c_i(k)$ represent, respectively, a global and an individual feedback signal whose purpose is to regulate the behaviour of each agent so as to achieve the desired level of compliance. Accordingly, we consider the following control laws, $\forall k \in \mathbb{N}$ and $\forall
 i \in \{1,\dots, n\}$:

\begin{equation}\label{con:GlobalCost}
C(k+1) = C(k) + \alpha (Q^* - n^{-1} \sum_{i=1}^n M_i(k - m)
\end{equation}

\begin{equation}\label{con:IndiCost}
c_i(k+1) = c_i(k) + \beta \, \left(Q^* - \overline{M_i(k - m)}\right)
\end{equation}
where $\alpha > 0$ and $\beta > 0$ are two constants, $\forall k \in \mathbb{N}$ and $\forall i \in \{1,\dots, n\}$, $Q^* \in [0,1]$ is the desired level of compliance,and $\overline{M_i(k)}$ is a windowed time average of the compliance of agent $i$, which is defined as

\begin{equation}
\overline{M_i(k)} = (1 - \gamma) \, \sum_{j=1}^k \gamma^{k-j} \, M_i(j).
\end{equation}
where $(1 - \gamma)^{-1}$ is the length of the window for the average, with  $\gamma < 1$.

Intuitively, this means that the value of the bond staked by agent $i$ depends on the overall compliance of all agents and on how agent $i$ behaved in the past. 
The use of both a global and an individual control signal brings several advantages, such as  \emph{Fairness}, \emph{Distributed trading of compliance levels} and resiliency from {\em Pricing attacks}. The interested reader can refer to \cite{ferraro2021personalised}.

\section{Mask model}
Given the general background described above, we consider scenarios where agents move within a confined and/or crowded space where the use of mask and social distancing norms is mandatory, such as buses, the tube or airplanes. 
The following steps are performed to encourage social compliance and the structure of the network is depicted in Figure \ref{fig: Network}. \newline

\begin{itemize}
\item When an agent wants to go into confined or crowded space, he must use a disposable mask with detachable integrated chip\footnote{The detachable integrated chip should include a main board, gas sensor for monitoring and battery for power. In this experiment, we set up a prototype: Raspberry Pi is adopted as the main board, USB connected to the computer to provide power. } connected to his phone or computer. This chip is connected to the customer's wallet.\newline
\item Each mask has a unique identifier, a detachable sensor and is connected to a Raspberry Pi which acts as the main computing unit (in a more realistic setting, the role of the Raspberry Pi would be carried out by other smaller computing units). \newline
\item Immediately after an agent enters a confined or crowded space, he has to deposit a stake a certain amount of tokens (determined by the PFCA), through a smart contract (this might be based on the past level of compliance of the agent and on the current average level of compliance ). \newline
\item The sensor chip monitors equivalent calculated Carbon-dioxide (eCO2) and Total Volatile Organic Compounds (TVOC), whereas the Raspberry Pi performs computations and uploads data to the IOTA Tangle in real time.\newline
\item If the customer wore the mask properly, they might either receive a portion of their tokens back or, depending on the specific policy employed (as mentioned in Section II.A), they have to stake more tokens and the process is repeated until they leave the confined or crowded space.\newline
\end{itemize}

We now provide a brief description about the hardware and software used for the smart mask prototype:\newline

\begin{itemize}
\item The Firefly Wallet\footnote{https://wiki.iota.org/learn/wallets/firefly-wallet, https://firefly.iota.org} is used as the customer's account and the tool to interact with the IOTA Tangle. \newline

\item A Raspberry Pi is used as the main board to collect sensor data and act as a node to send and publish data to IOTA Tangle.
\newline

\item A gas sensor, CJMCU-811 is used to monitor the density of eCO2 and TVOC that people exhale.\newline

\end{itemize}

 \textbf{Remark:}  As this is a PoC, we use a Raspberry Pi as part of the prototype mainly for convenience. In an actual implementation, one would use a small chip integrated into the mask for detecting data (to be mounted on the mask) and another computing units (such as a phone, a computer, etc.) for collecting and analysing data. 



\begin{figure*}
\centering
\includegraphics[width=15cm, height=6.7cm]{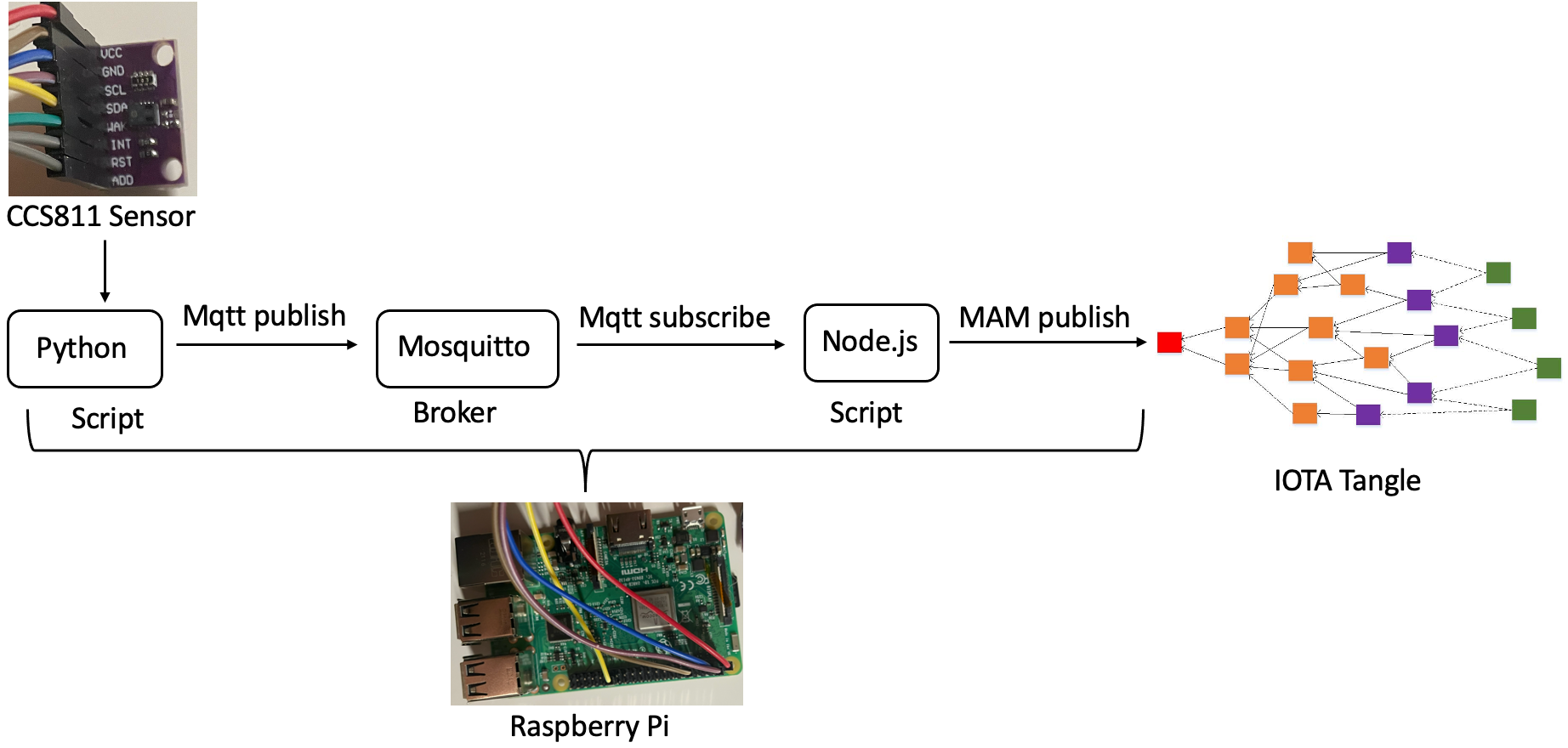}
\caption{ Implementation Architecture}
\label{fig: Netstruc}
\end{figure*}

\subsection{The Communication Protocol}

The network flow to upload data to the IOTA Tangle is depicted in Figure \ref{fig: Netstruc}.
The data is collected from the CJMCU-811 sensor which is connected to the Raspberry Pi 3B. As the most common communication protocol in IoT systems, a lightweight open source Message Queuing Telemetry Transport (MQTT) \cite{al2020investigating,silva2018latency} is adopted to transfer data in the network. 
MQTT works in a publish and subscribe model, in which some devices publish messages on a topic, while some devices which have subscribed to this topic receive this message.
After the Raspberry Pi gathers eCo2  and TVOC data from the sensor, it publishes these data to a specific topic\footnote{Other IoT devices such as ESP32, ESP8266, and Arduino can also be used to collect data and publish them through MQTT protocol. Moreover, more sensors, like temperature and humidity sensor can also be added to monitor customers' status.} \cite{strugar2018m2m}. 
The module depicted in Figure \ref{fig: Netstruc} named Node.js is set as the back-end server to subscribe to this specific topic and publish contained data in this topic to the IOTA Tangle though a lightweight data transmission protocol – Masked Authenticated Messaging (MAM) \cite{brogan2018authenticating, zheng2019accelerating}.\newline

 The MAM is an important method for securing the transfer or access of the data stored in the IOTA Tangle. Nodes or devices, which are connected to the IOTA Tangle acting as publishers, broadcast their encrypted messages into a specified channel, while nodes which are interested in receiving the published messages can subscribe to the same channel \cite{brogan2018authenticating}. There are three modes within MAM, including Public mode, Private mode, Restricted mode. For public MAM mode, it employs the address of transaction which is same as Merkle Tree's root \cite{florea2018blockchain}. For private MAM mode, it employs the addresses of transactions which is gained by hashing the root of the Merkle Tree, which means the message can be known only if the root of the Merkle Tree is obtained. For restricted MAM mode, it employs the address of transaction which is obtained by the hash of the root of the Merkle tree and a side key, which means the message can be known only if both the key and the root are known \cite{bhandary2020blockchain}. \newline 
 


\section{Experiment set up}

\begin{figure}
\centering
\includegraphics[width=0.97\columnwidth]{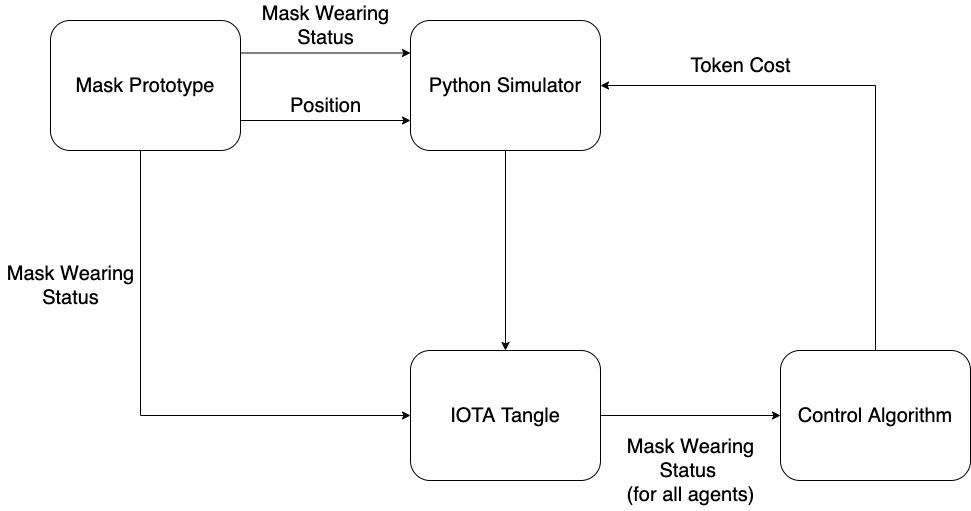}
\caption{Simulation structure}
\label{fig: simula_struc}
\end{figure}

To showcase the functioning of the mask prototype we set up  a hardware-in-the-loop kind of simulation based on the structure depicted in Figure \ref{fig: simula_struc}. The simulation is divided into the following components:

\begin{itemize}
    \item \emph{Agent-based Simulator:} an agent-based simulation is employed to mimic the presence of multiple agent within a room. Each agent behaves according to the equations described in Section II.\newline
    
    \item \emph{Smark mask prototype:} a user wears the mask prototype and its position and mask wearing status are respectively recorded to the IOTA Tangle and sent to the agent-based Simulator in real time, as if the user was one of the agents of the simulations.\newline
    
    \item \emph{Controller:} the controller uses the data stored in the IOTA Tangle to compute in real time the individual cost for each agent, according to Equations \ref{con:GlobalCost} and \ref{con:IndiCost}.\newline
\end{itemize}

In what follows, we provide a detailed description of each component used to perform this simulation.

\subsection{Python-based simulation}
To illustrate the effectiveness of wearing masks in confined and crowded spaces, we perform simulations in a similar way to the ones in \cite{ferraro2021personalised}. We make use of an agent-based model to simulate the spread of covid-19 within a confined space. In this experiment, we assume the number of agents is 500 and the age of these people is following a normal distribution\footnote{https://www.trustforlondon.org.uk/data/population-age-groups/}. Whenever agent $i$ gets in close range with agent $j$, i.e. when $||x_i - x_j||_2 \leq \epsilon$, there is a positive chance that agent $i$ might get infected (given that agent $j$ is positive). This chance is defined as
\begin{equation}
  P(\text{$i$ get infected}) = P_0*(1-m_i*M_i(k))*(1-m_j*M_j(k))
 \label{eq: prob}
\end{equation}

where the base infection rate is denoted by $P_0$ (the chance of getting infected with no social distancing and mask wearing),  $m_i \in [0,1]$ is the effectiveness of the mask worn by agent $i$ and $M_i(k) \in \{0,1\}$ is a binary random variable that indicates whether agent $i$ is wearing a mask at time $k$. \newline

Simulations show the following results:\newline

\begin{figure*}
	\centering
	     \subfigure[t =10 seconds]{
    	\begin{minipage}[b]{0.2\linewidth}
   		\includegraphics[width=4cm, height=3.3cm]{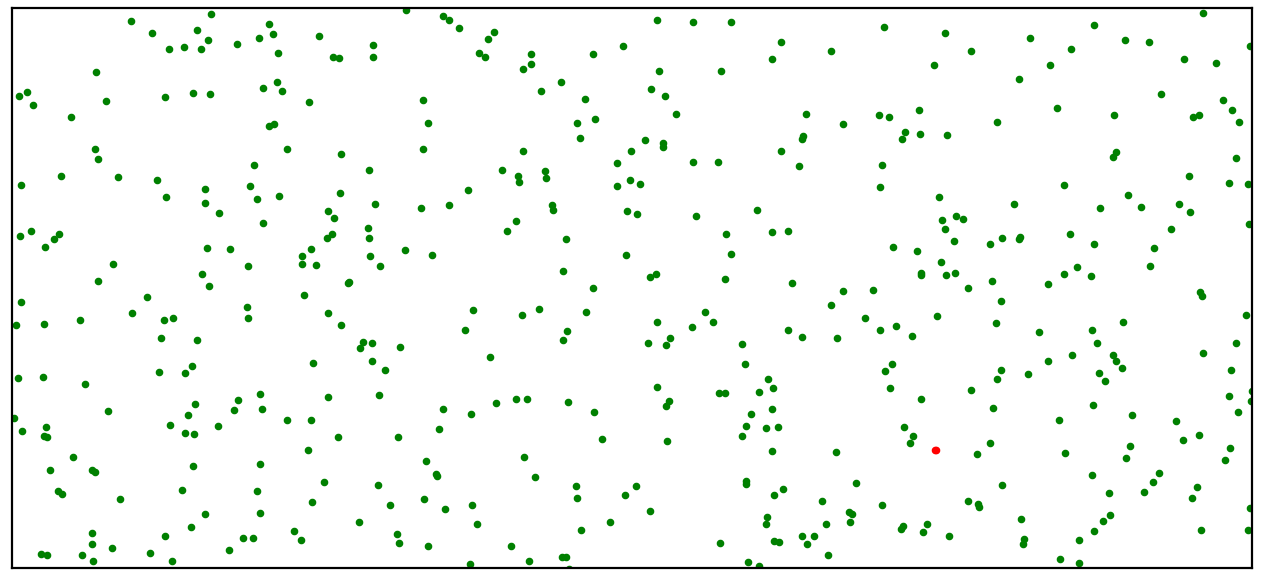}
    	\end{minipage}
	\label{fig:Without10}
    }
    \hspace{0.1mm}
    \subfigure[t =15 seconds]{
    	\begin{minipage}[b]{0.2\linewidth}
   		\includegraphics[width=4cm, height=3.3cm]{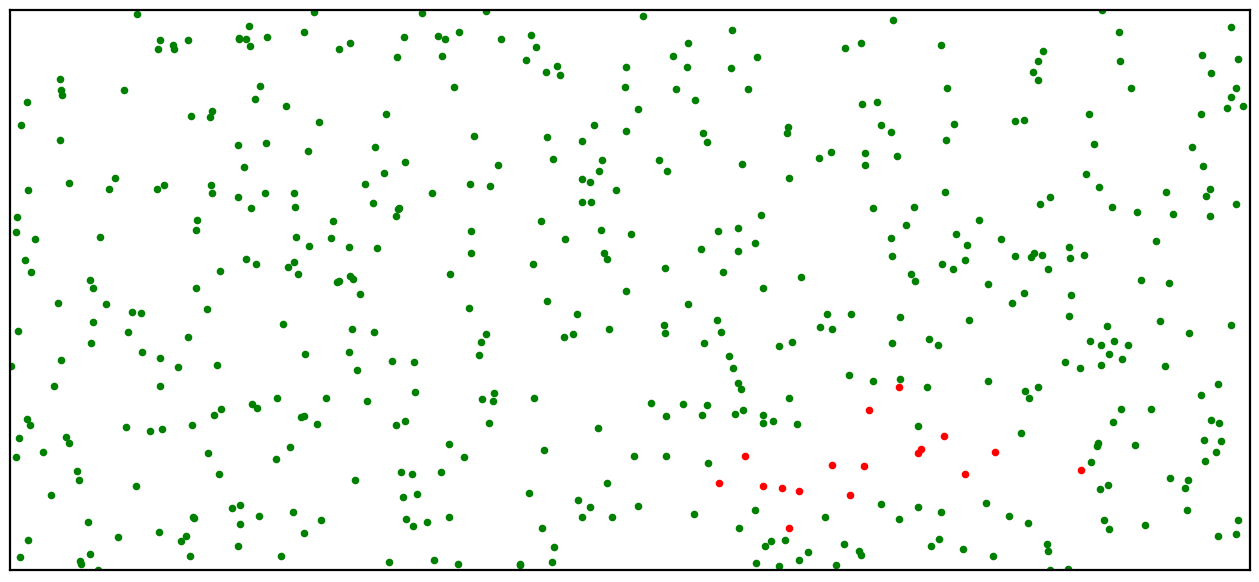}
    	\end{minipage}
	\label{fig:Without15}
    }
	\hspace{0.1mm}
		\subfigure[t =20 seconds]{
		\begin{minipage}[b]{0.2\linewidth}
			\includegraphics[width=4cm, height=3.3cm]{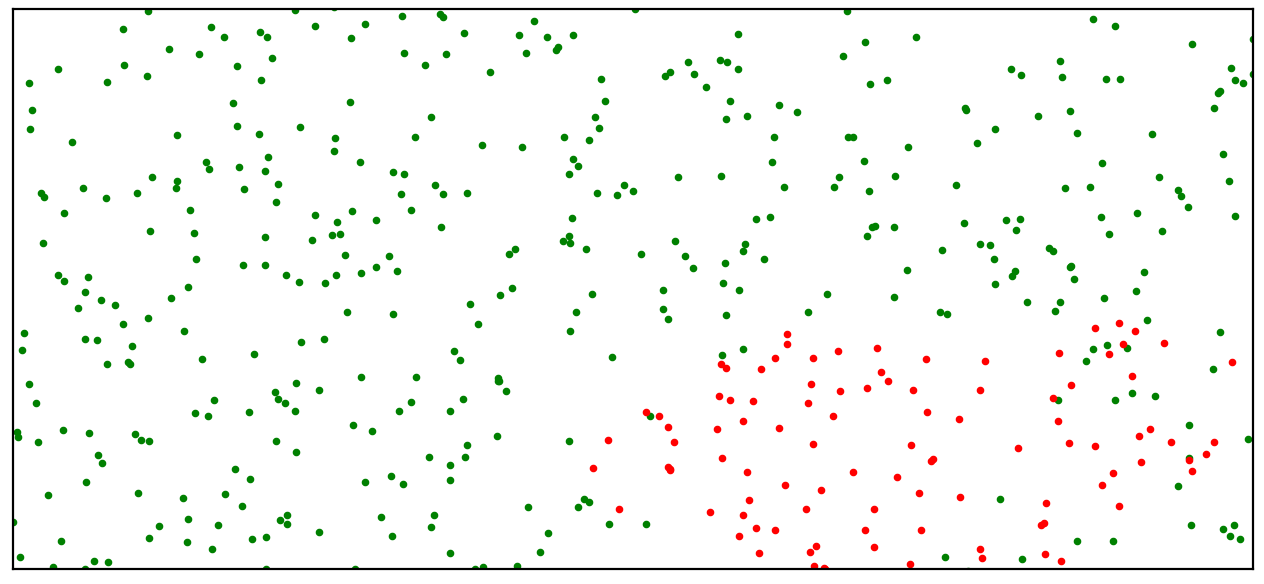}
		\end{minipage}
		\label{fig:Without20}
	}
	\hspace{0.1mm}
		\subfigure[t =25 seconds]{
		\begin{minipage}[b]{0.2\linewidth}
			\includegraphics[width=4cm, height=3.3cm]{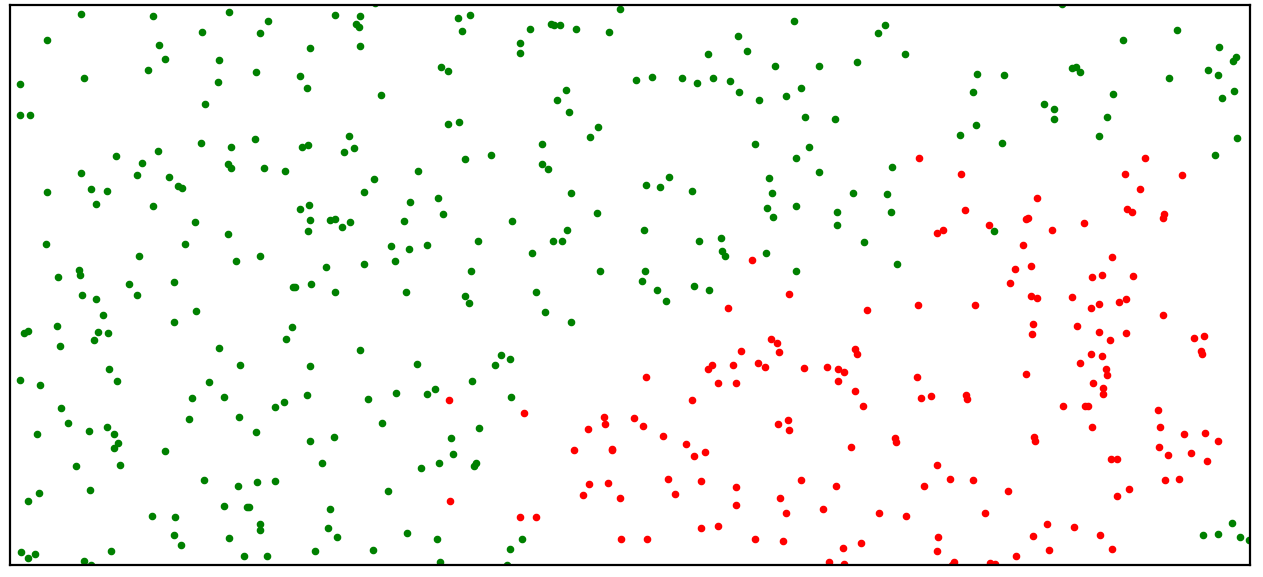}
		\end{minipage}
		\label{fig:Without25}
	}
	\hspace{0.1mm}
    	\subfigure[t =30 seconds]{
		\begin{minipage}[b]{0.2\linewidth}
			\includegraphics[width=4cm, height=3.3cm]{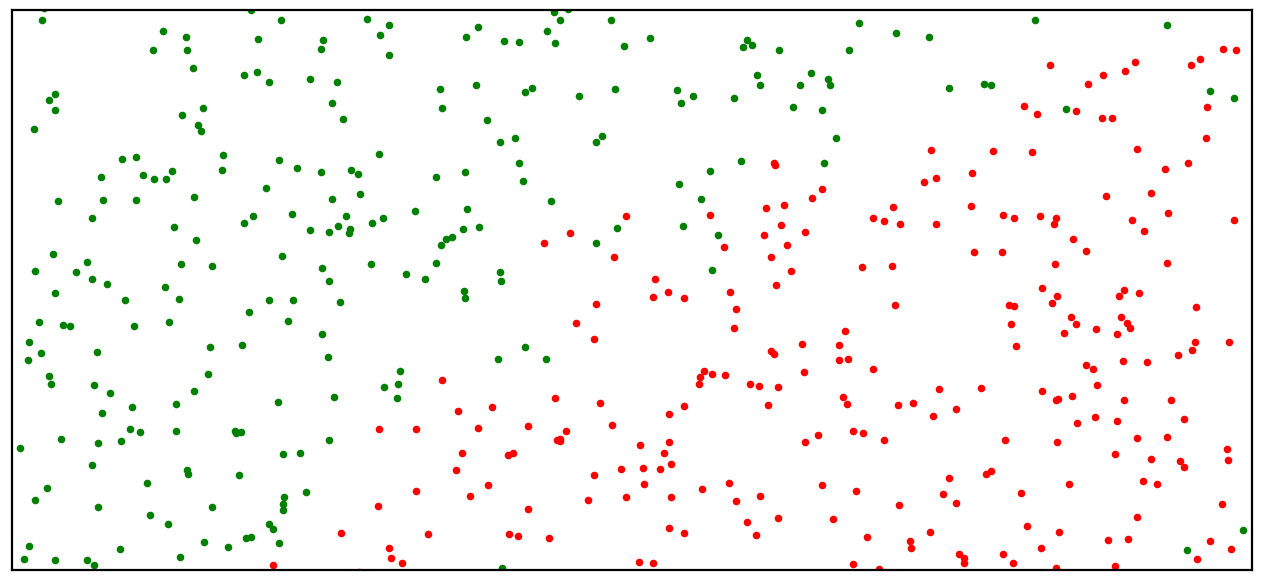}
		\end{minipage}
		\label{fig:Without30}
}
\hspace{0.1mm}
    \subfigure[t =35 seconds]{
    	\begin{minipage}[b]{0.2\linewidth}
   		\includegraphics[width=4cm, height=3.3cm]{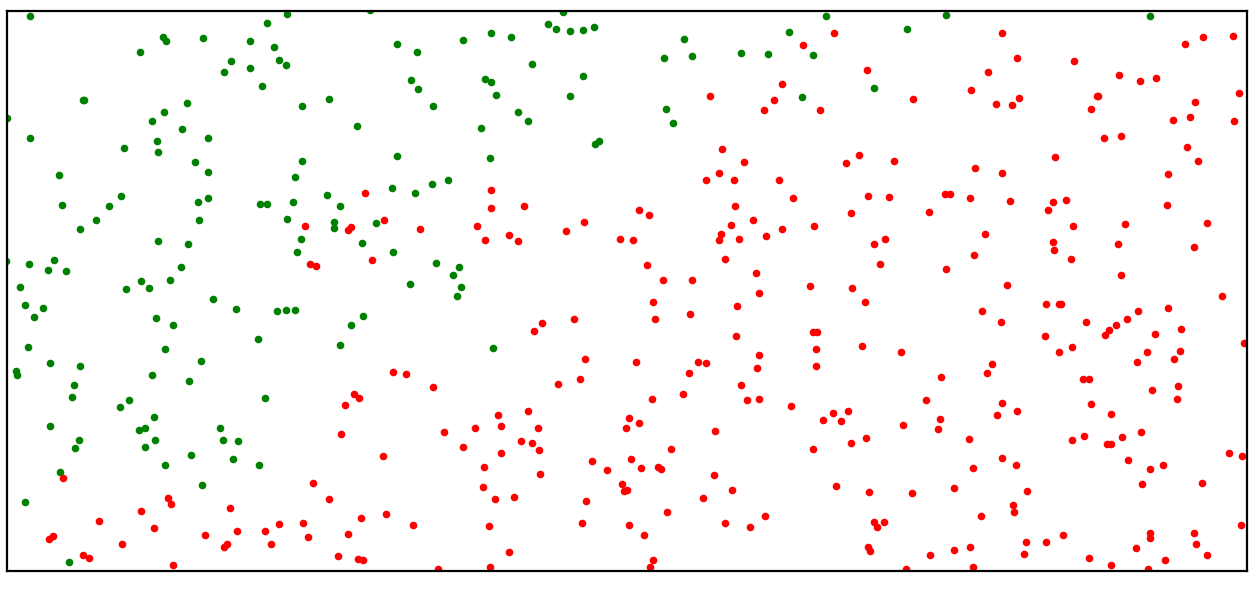}
    	\end{minipage}
	\label{fig:Without35}
    }
\hspace{0.1mm}
    \subfigure[t =45 seconds]{
    	\begin{minipage}[b]{0.2\linewidth}
   		\includegraphics[width=4cm, height=3.3cm]{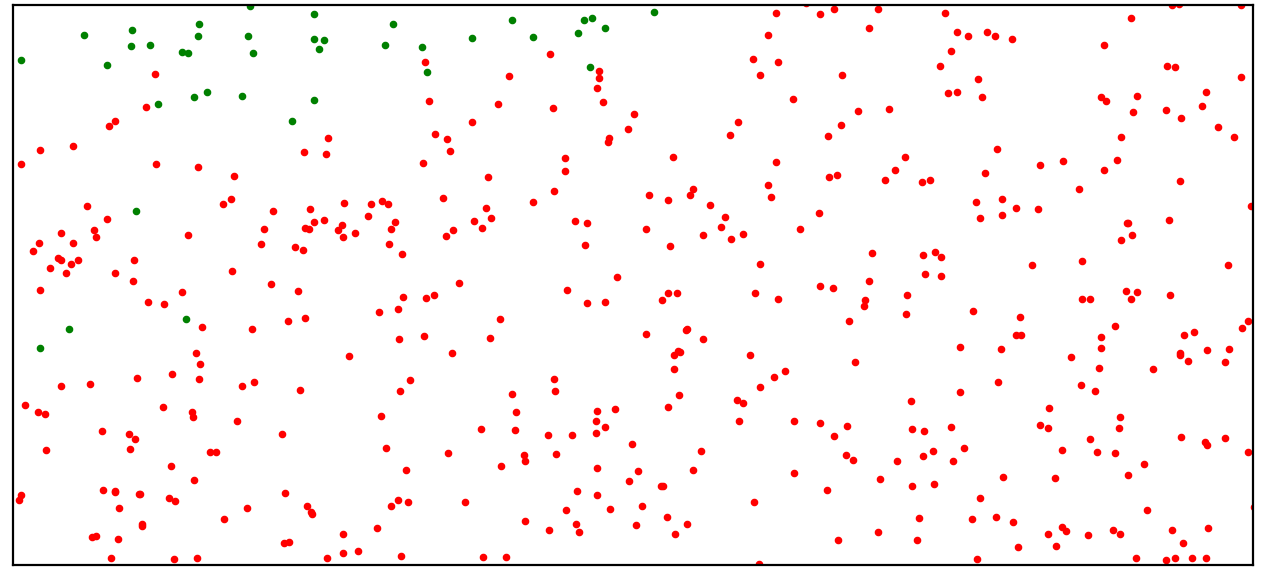}
    	\end{minipage}
	\label{fig:Without50}
    }    
    \hspace{0.1mm}
    \subfigure[t =55 seconds]{
    	\begin{minipage}[b]{0.2\linewidth}
   		\includegraphics[width=4cm, height=3.3cm]{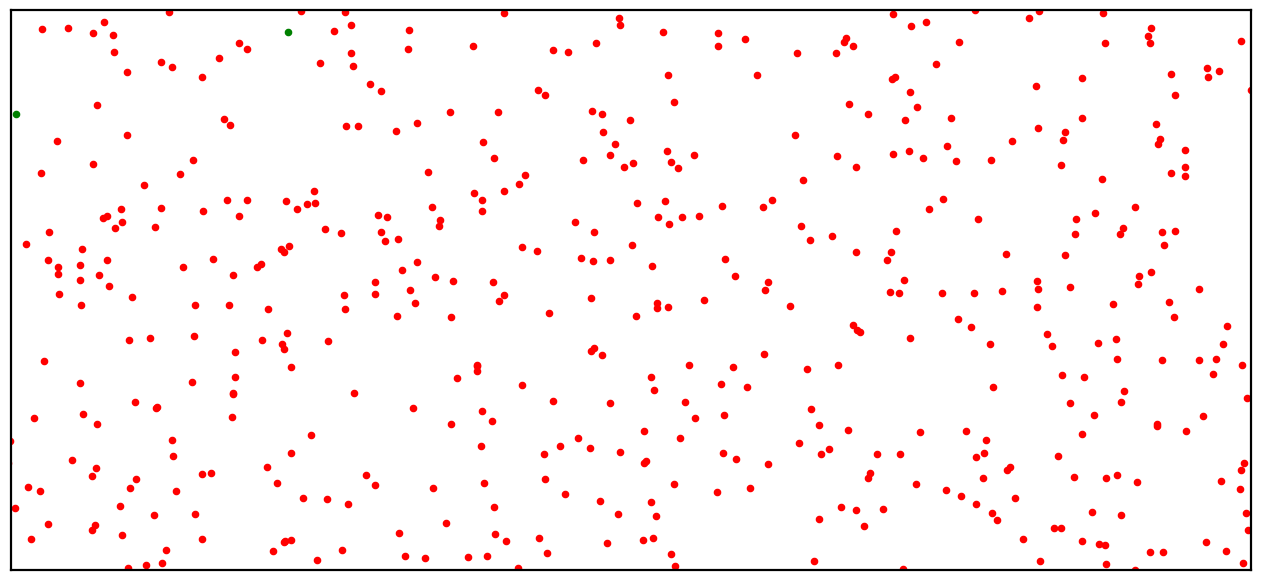}
    	\end{minipage}
	\label{fig:Without55}
    }    
\hspace{0.1mm}
    \subfigure[t =87 seconds]{
    	\begin{minipage}[b]{0.2\linewidth}
   		\includegraphics[width=4cm, height=3.3cm]{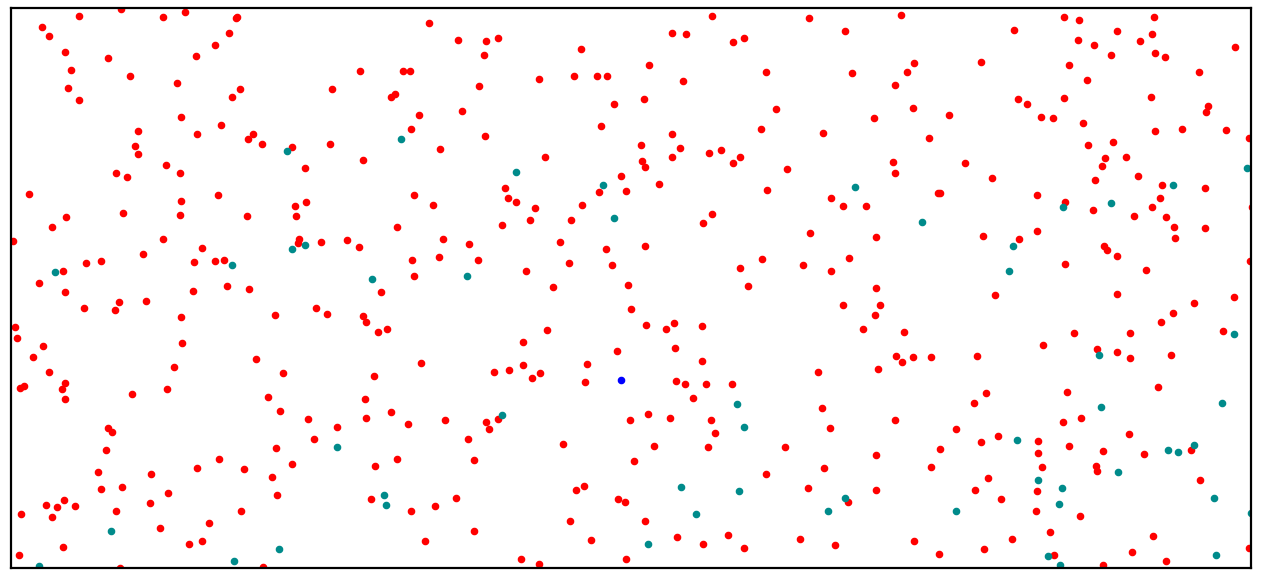}
    	\end{minipage}
	\label{fig:Without87}
    }
\hspace{0.1mm}
    \subfigure[t =108 seconds]{
    	\begin{minipage}[b]{0.2\linewidth}
   		\includegraphics[width=4cm, height=3.3cm]{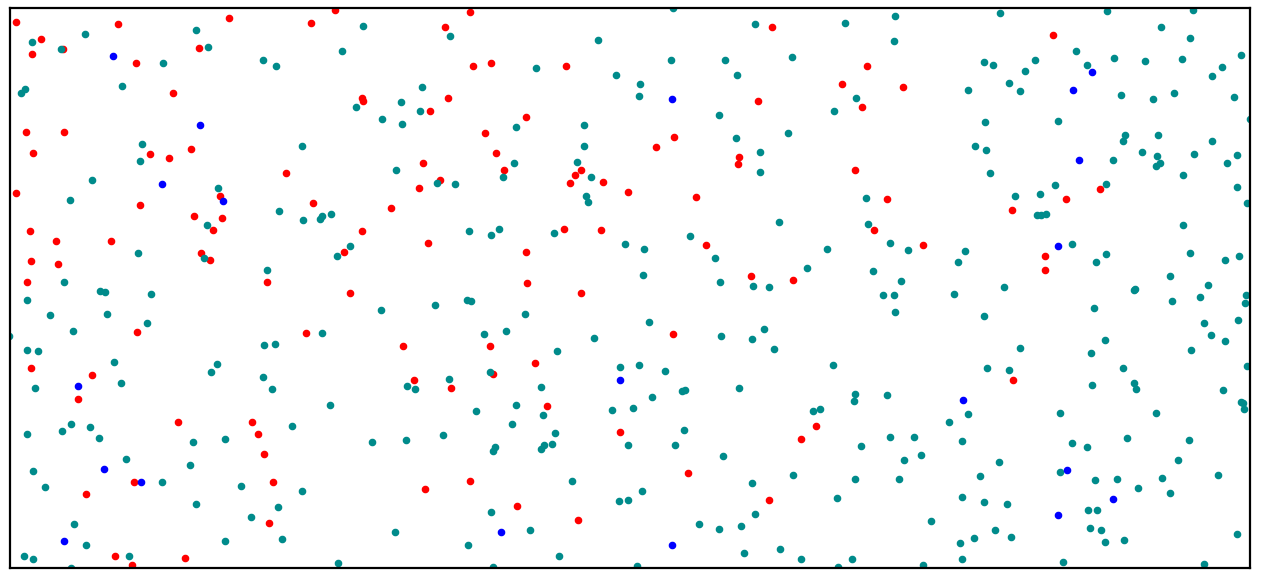}
    	\end{minipage}
	\label{fig:Without108}
    }    
    \hspace{0.1mm}
    \subfigure[t =108 seconds]{
    	\begin{minipage}[b]{0.2\linewidth}
   		\includegraphics[width=4cm, height=3.3cm]{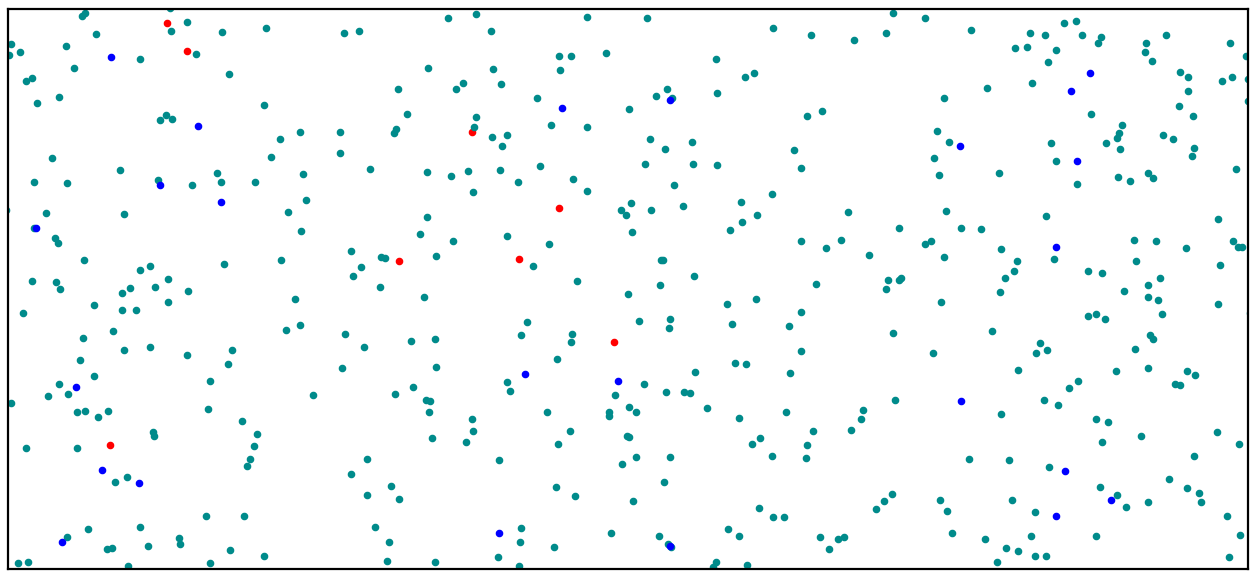}
    	\end{minipage}
	\label{fig:Without122}
    }      
        \hspace{0.1mm}
    \subfigure[t =133 seconds]{
    	\begin{minipage}[b]{0.2\linewidth}
   		\includegraphics[width=4cm, height=3.3cm]{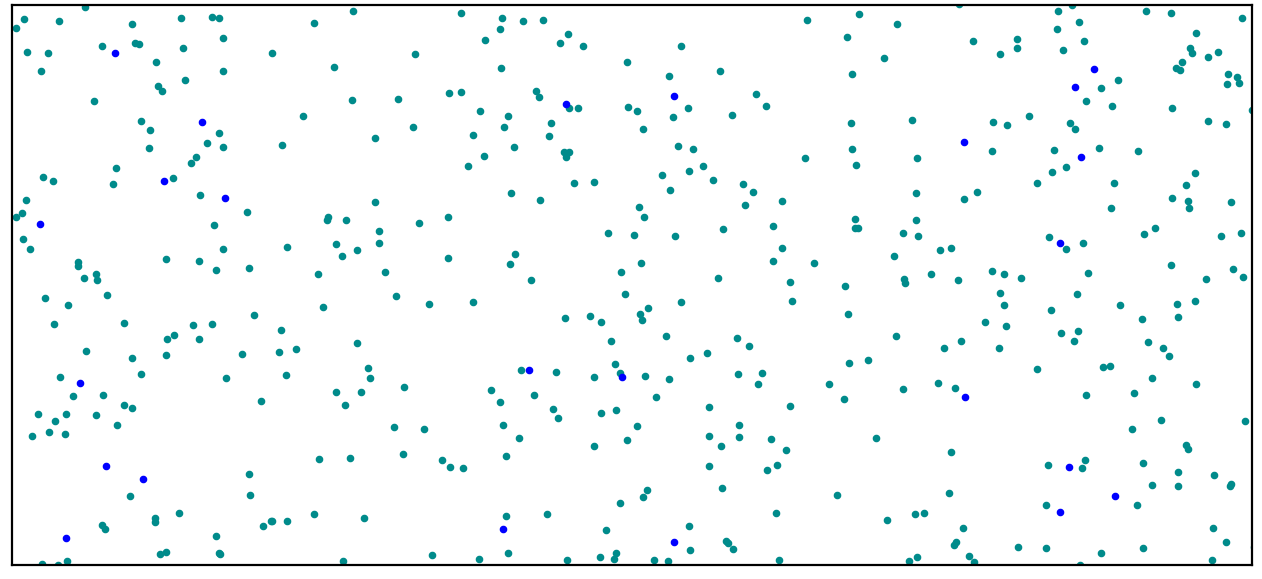}
    	\end{minipage}
	\label{fig:Without133}
    }        
	\caption{The infection among people when they do not wear masks properly.  Green dots denote healthy people, red dots denote infected people, cyan dots denote immune people with slight possible sequelae, while blue dots denote immune people with serious sequelae.}
	\label{fig: Nomask}
\end{figure*}

\begin{figure*}
	\centering
	     \subfigure[20\%-30\% mask wearing ]{
    	\begin{minipage}[b]{0.3\linewidth}
   		\includegraphics[width=6.5cm, height=4cm]{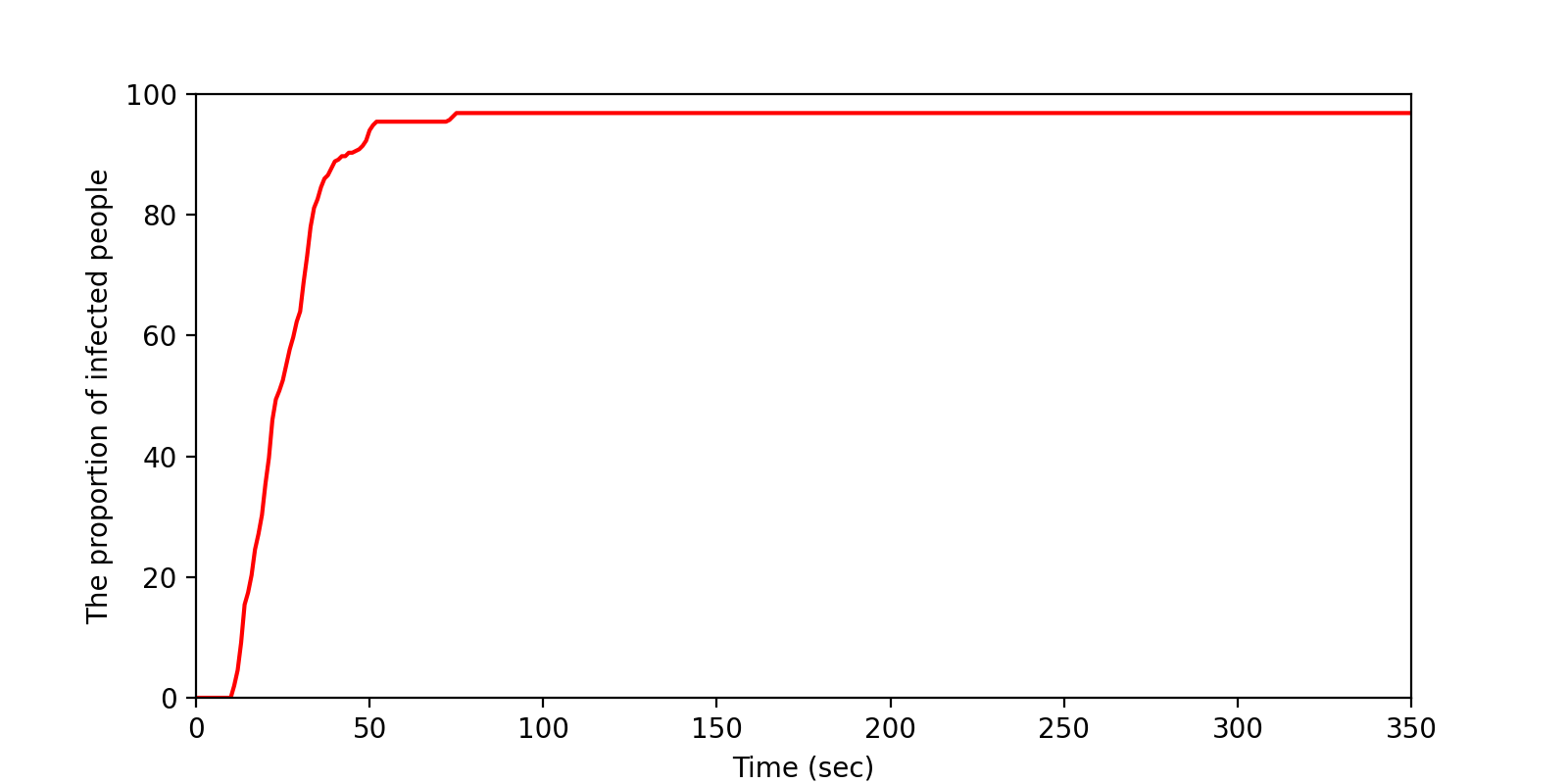}
    	\end{minipage}
	\label{fig:02mask}
    }
    \hspace{1mm}
	\subfigure[30\%-40\% mask wearing ]{
    	\begin{minipage}[b]{0.3\linewidth}
   		\includegraphics[width=6.5cm, height=4cm]{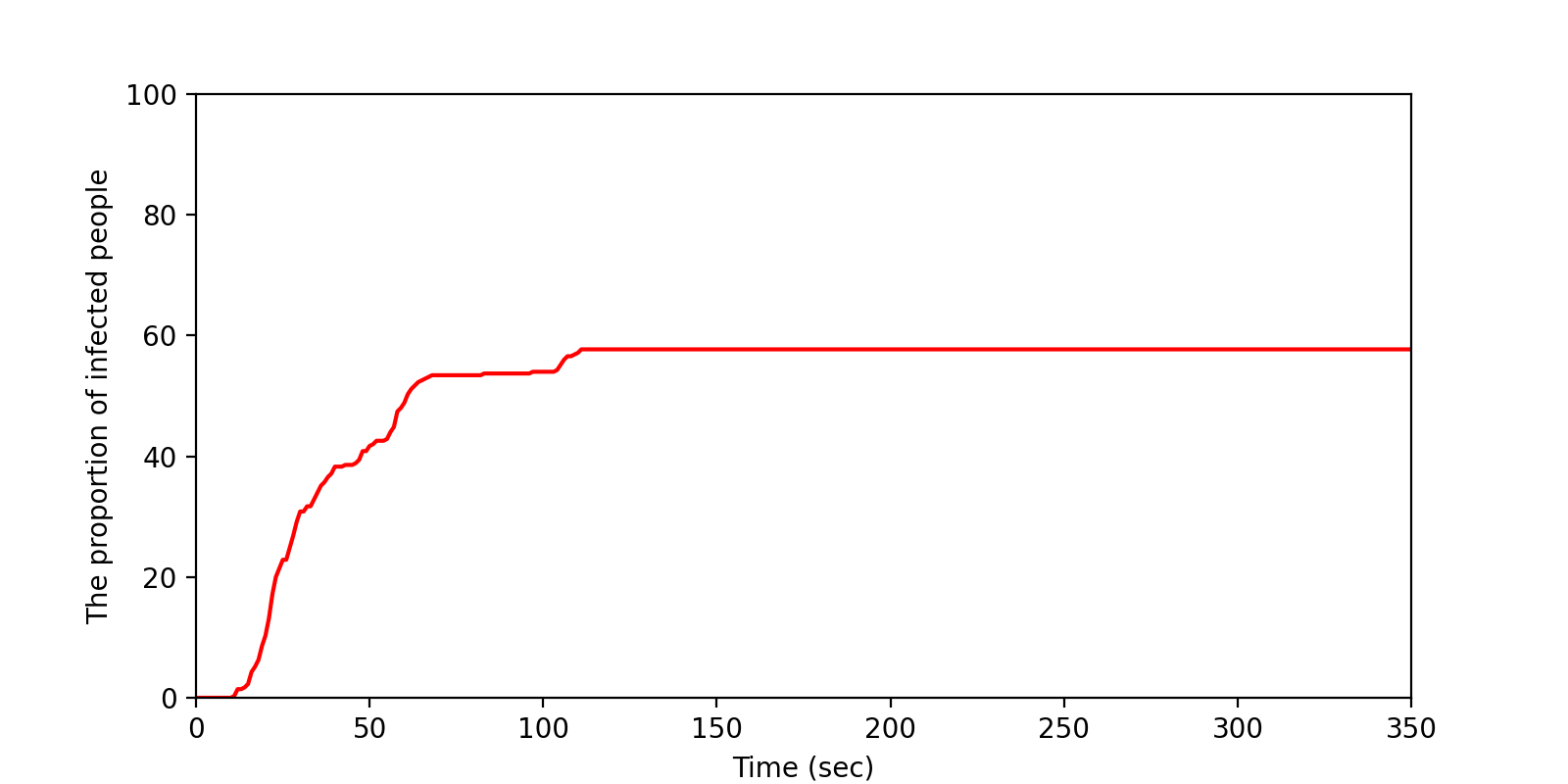}
    	\end{minipage}
	\label{fig:03mask}
    }
    \hspace{1mm}
    \subfigure[50\%-60\% mask wearing ]{
    	\begin{minipage}[b]{0.3\linewidth}
   		\includegraphics[width=6.5cm, height=4cm]{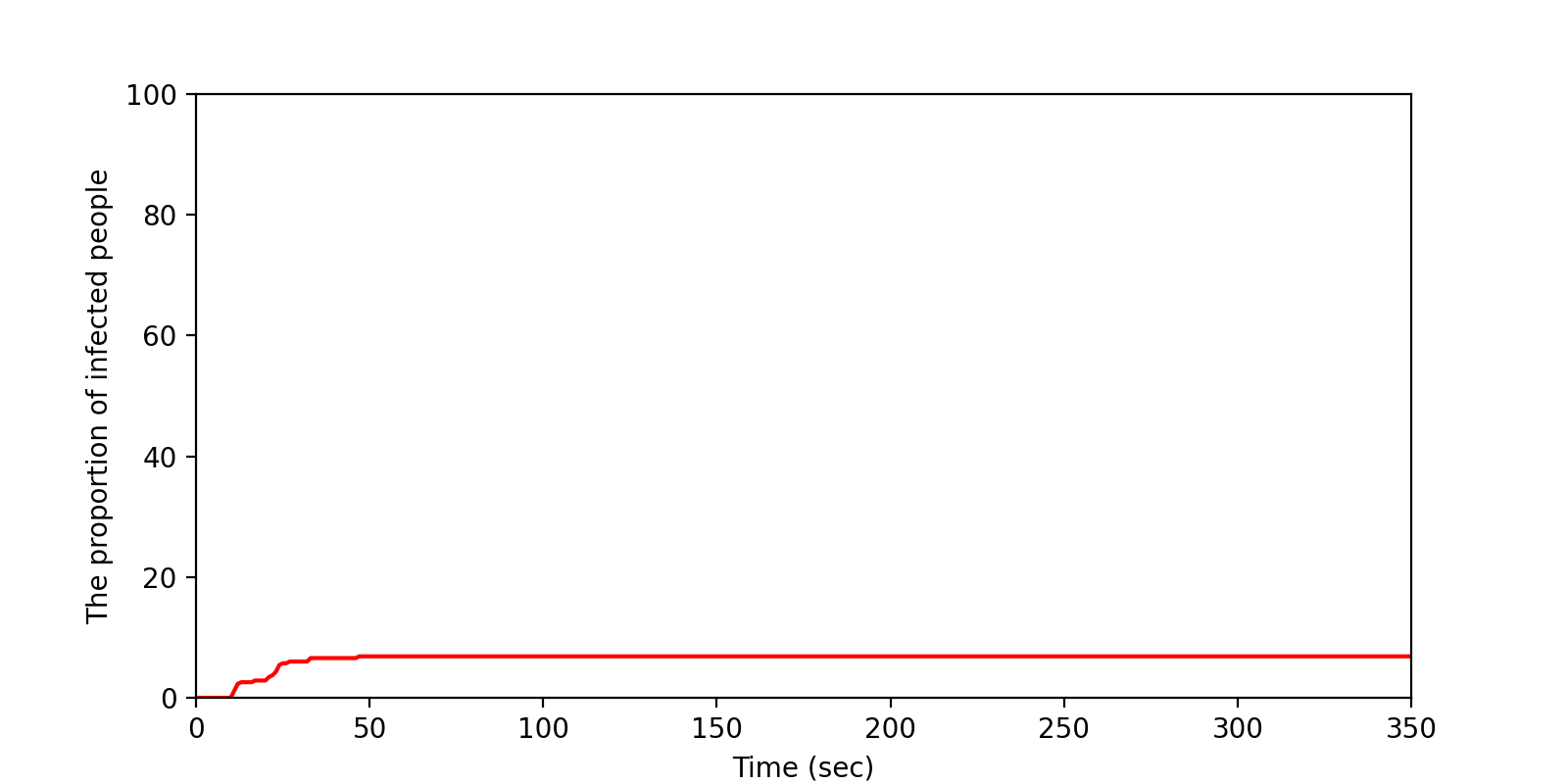}
    	\end{minipage}
	\label{fig:05mask}
    }
	\caption{The number of infected people}
	\label{fig: Trend}
\end{figure*}

\begin{itemize}

\item  When no masks are worn the infection rate increases very quickly. As shown in Figure \ref{fig:Without10}, we set up the model with a single infected individual, marked as the red dot. As depicted in Figures \ref{fig:Without15}-\ref{fig:Without55} virus spreads very quickly: at time 30, depicted in Figure \ref{fig:Without30}, nearly half people get infected and nearly all agents get infected at time 55 as depicted in Figure \ref{fig:Without55}.\newline

\item When people, including healthy people and infected people wear masks effectively, the infection rate and serious infection rate increases at a slower rate. As depicted in Figure \ref{fig:02mask}, when 20\%-30\% people wear masks, the proportion of infected people is almost 100\% around 50 seconds; As depicted in Figure \ref{fig:03mask}, when 30\%-40\% people wearing masks, the maximum percentage people get infected is also around 60\%; While as depicted in \ref{fig:05mask}, when 50\%-60\% people are wearing masks, the maximum percentage of infected people is only around 10\% at time 50. Finally, Figure \ref{fig:07mask} shows that the amount of infected people is even lower if 70\%-80\% people are wearing masks.\newline
\end{itemize}

\begin{figure}
\centering
\includegraphics[width=0.97\columnwidth]{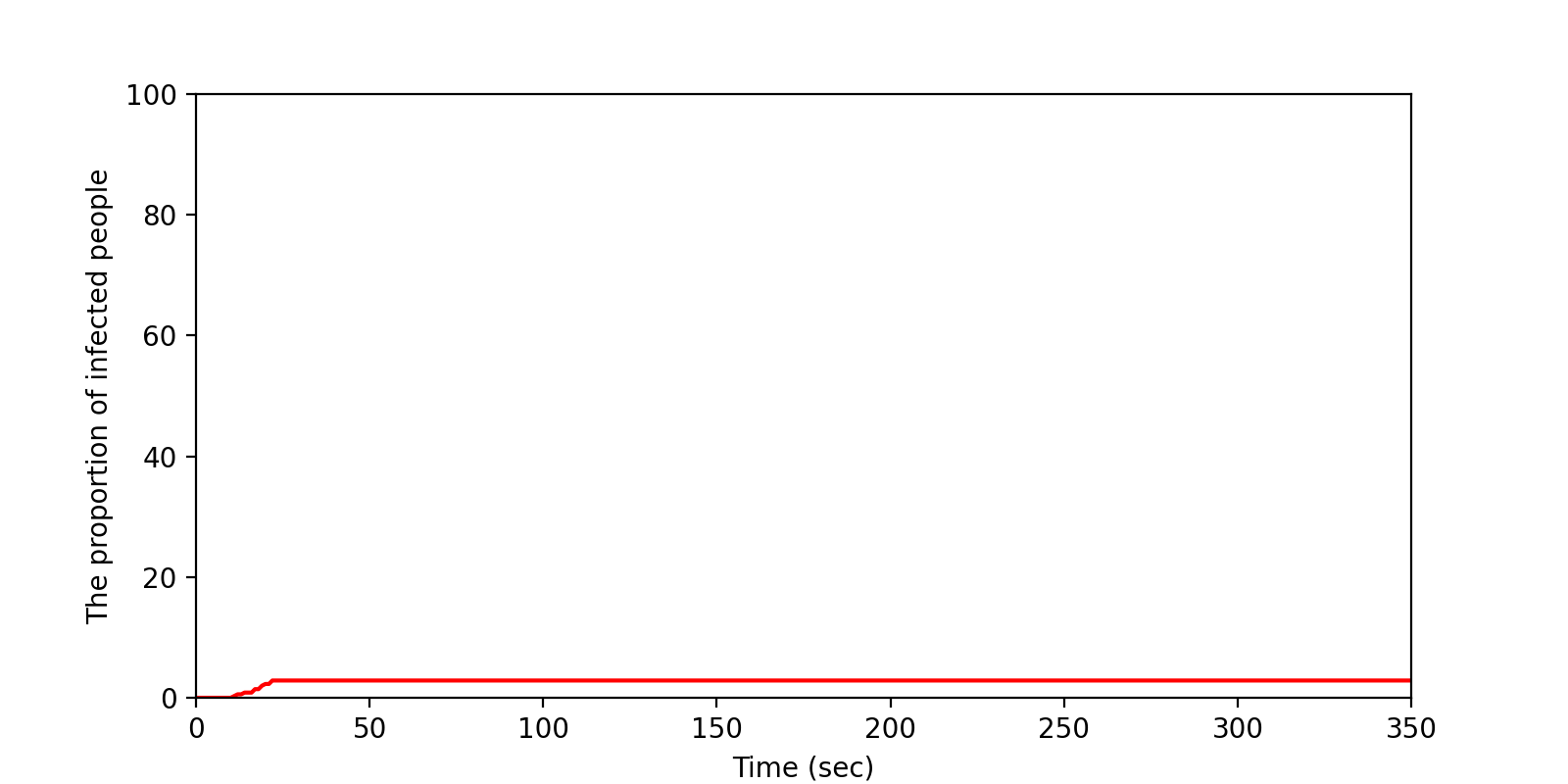}
\caption{70\%-80\% mask wearing}
\label{fig: 07mask}
\end{figure}

 \begin{figure*}
	\centering
	     \subfigure[ Cost]{
    	\begin{minipage}[b]{0.45\linewidth}
   		\includegraphics[width=9.7cm, height=8.5cm]{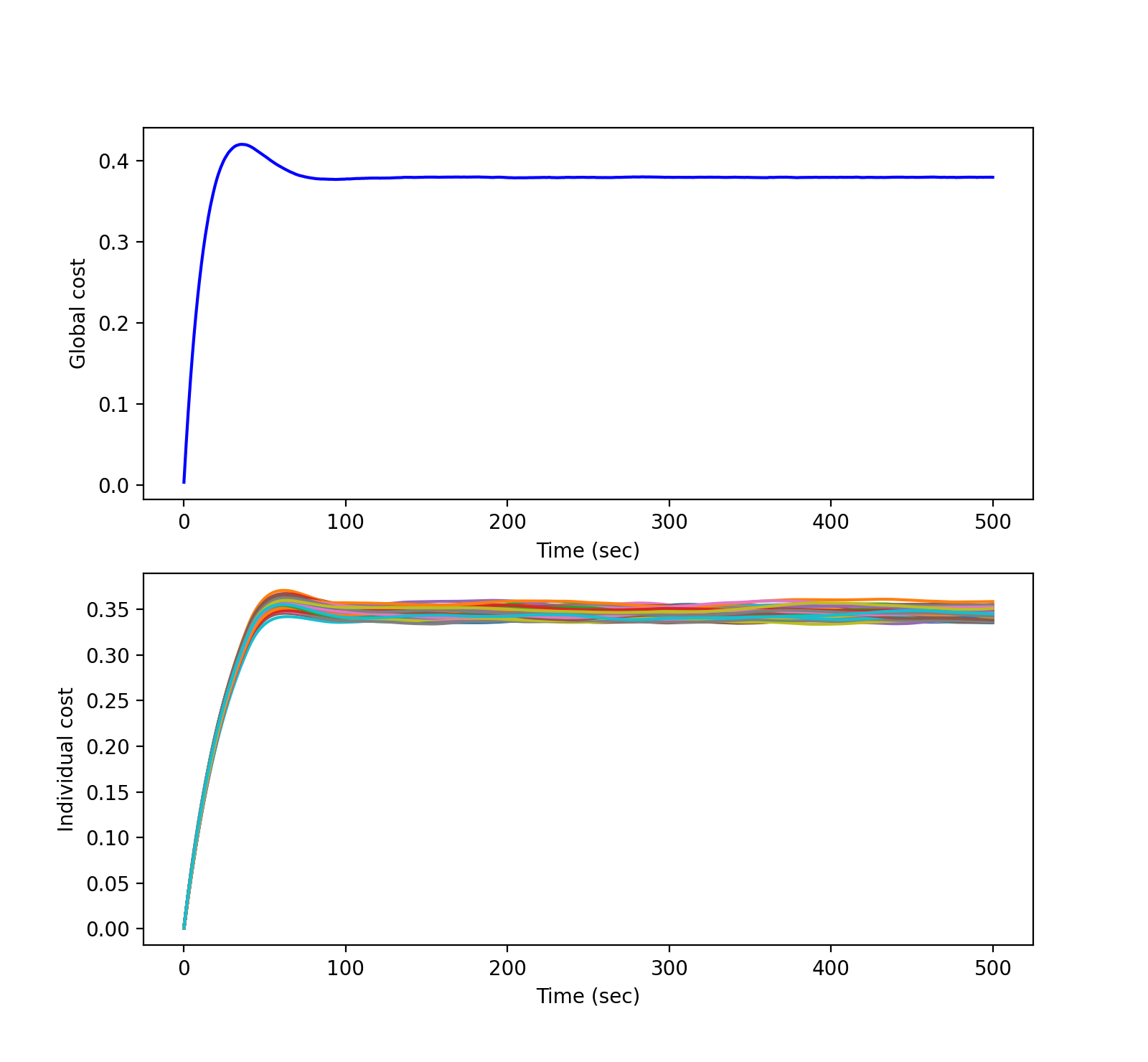}
    	\end{minipage}
	\label{fig:Cost}
    }
    \hspace{3mm}
    \subfigure[Compliance ]{
    	\begin{minipage}[b]{0.45\linewidth}
   		\includegraphics[width=9.7cm, height=8.5cm]{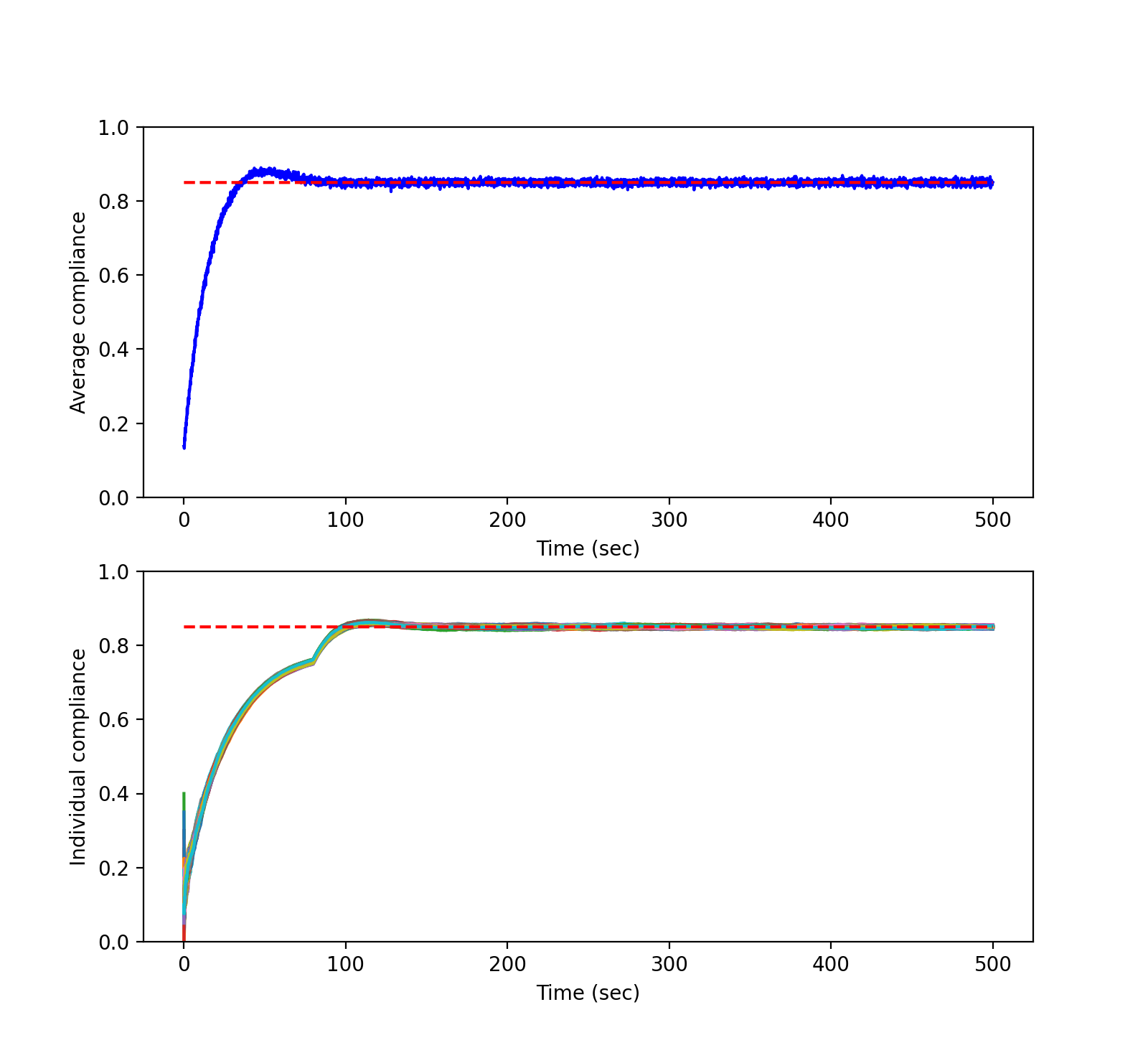}
    	\end{minipage}
	\label{fig:Compliance}
    }
	\caption{Compliance control with both global and the individual signals.}
	\label{fig: Cost_Compliance}	
\end{figure*}

 \begin{figure*}
	\centering
	     \subfigure[ Mask wearing test]{
    	\begin{minipage}[b]{0.45\linewidth}
   		\includegraphics[width=9cm, height=4.8cm]{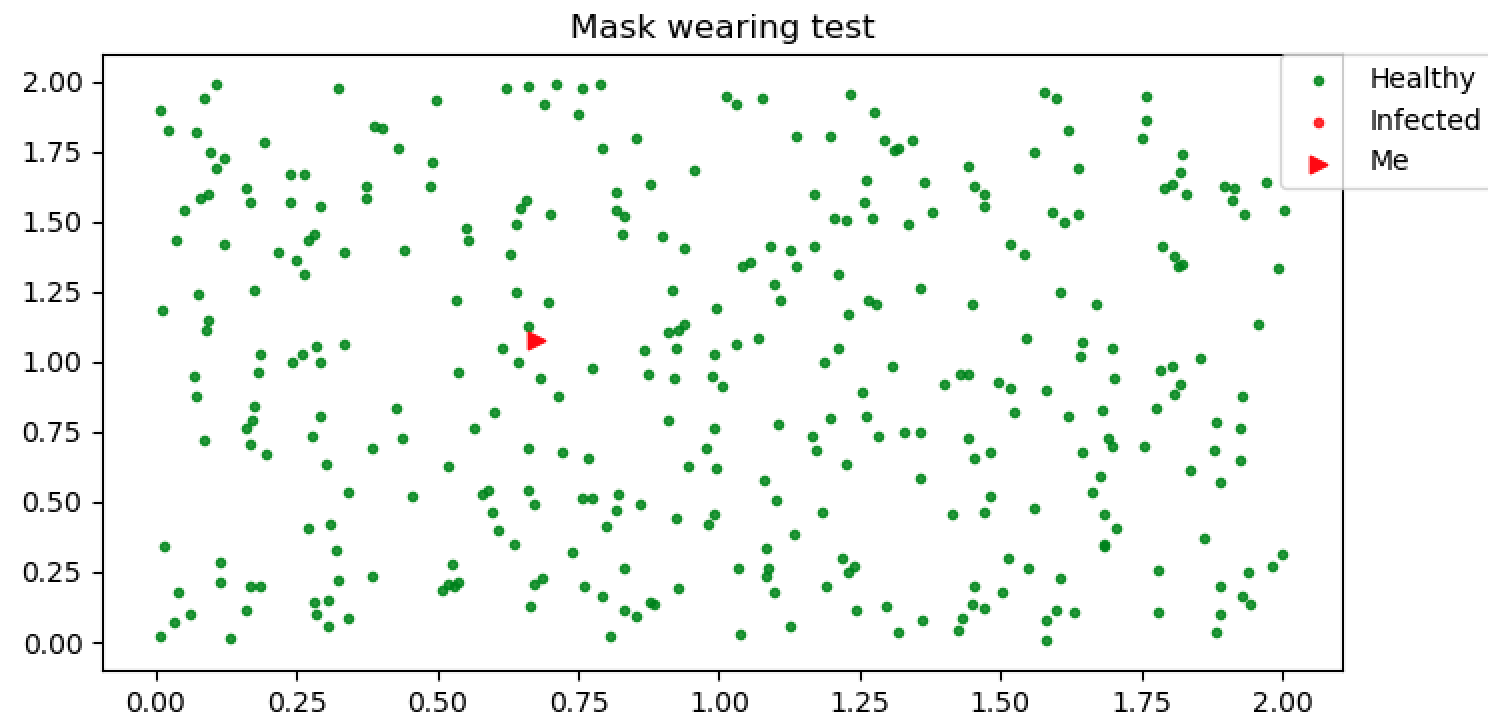}
    	\end{minipage}
	\label{fig:Realsim1}
    }
    \hspace{3mm}
    \subfigure[Mask wearing test ]{
    	\begin{minipage}[b]{0.45\linewidth}
   		\includegraphics[width=9cm, height=4.8cm]{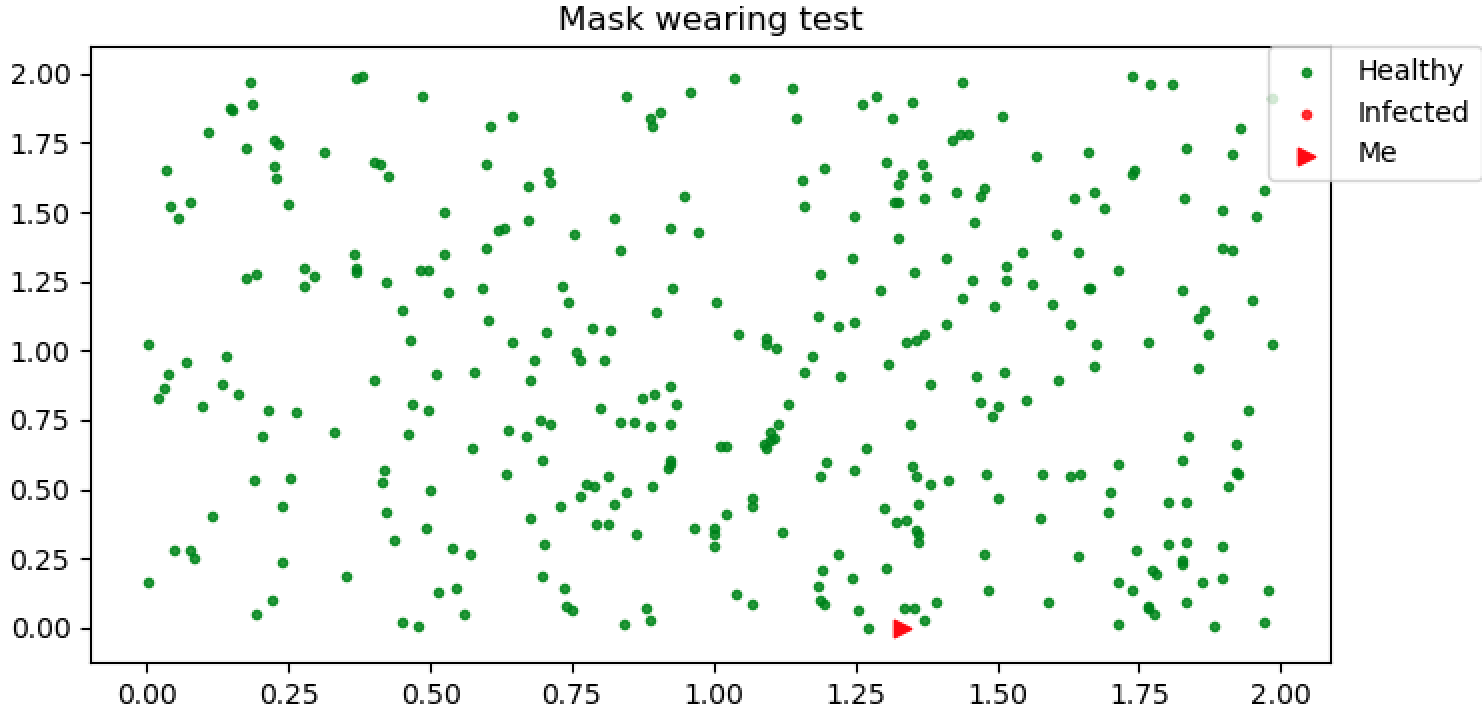}
    	\end{minipage}
	\label{fig:Realsim2}
    }
    \hspace{3mm}
    \subfigure[Co2 and TVOC value ]{
    	\begin{minipage}[b]{0.45\linewidth}
   		\includegraphics[width=8cm, height=4.8cm]{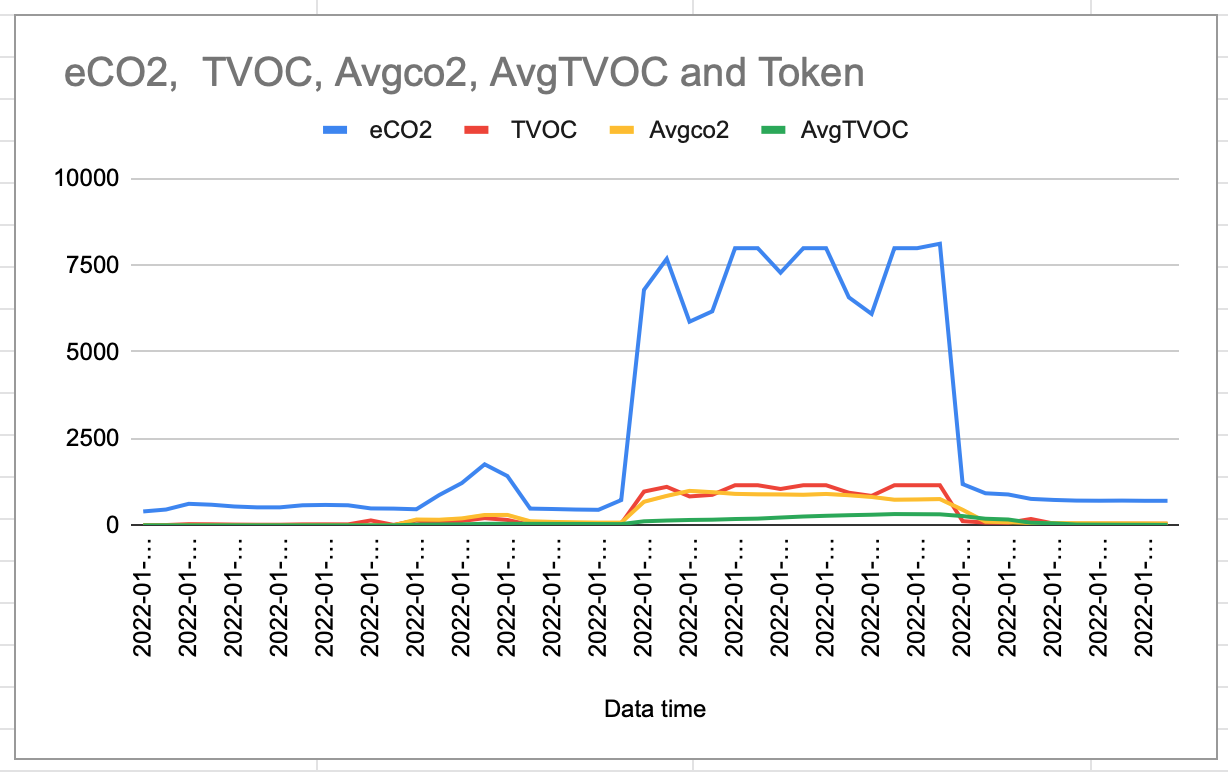}
    	\end{minipage}
	\label{fig:co2tvoc}
    }
    \hspace{-6mm}
    \subfigure[Token's value ]{
    	\begin{minipage}[b]{0.45\linewidth}
   		\includegraphics[width=9.5cm, height=5.3cm]{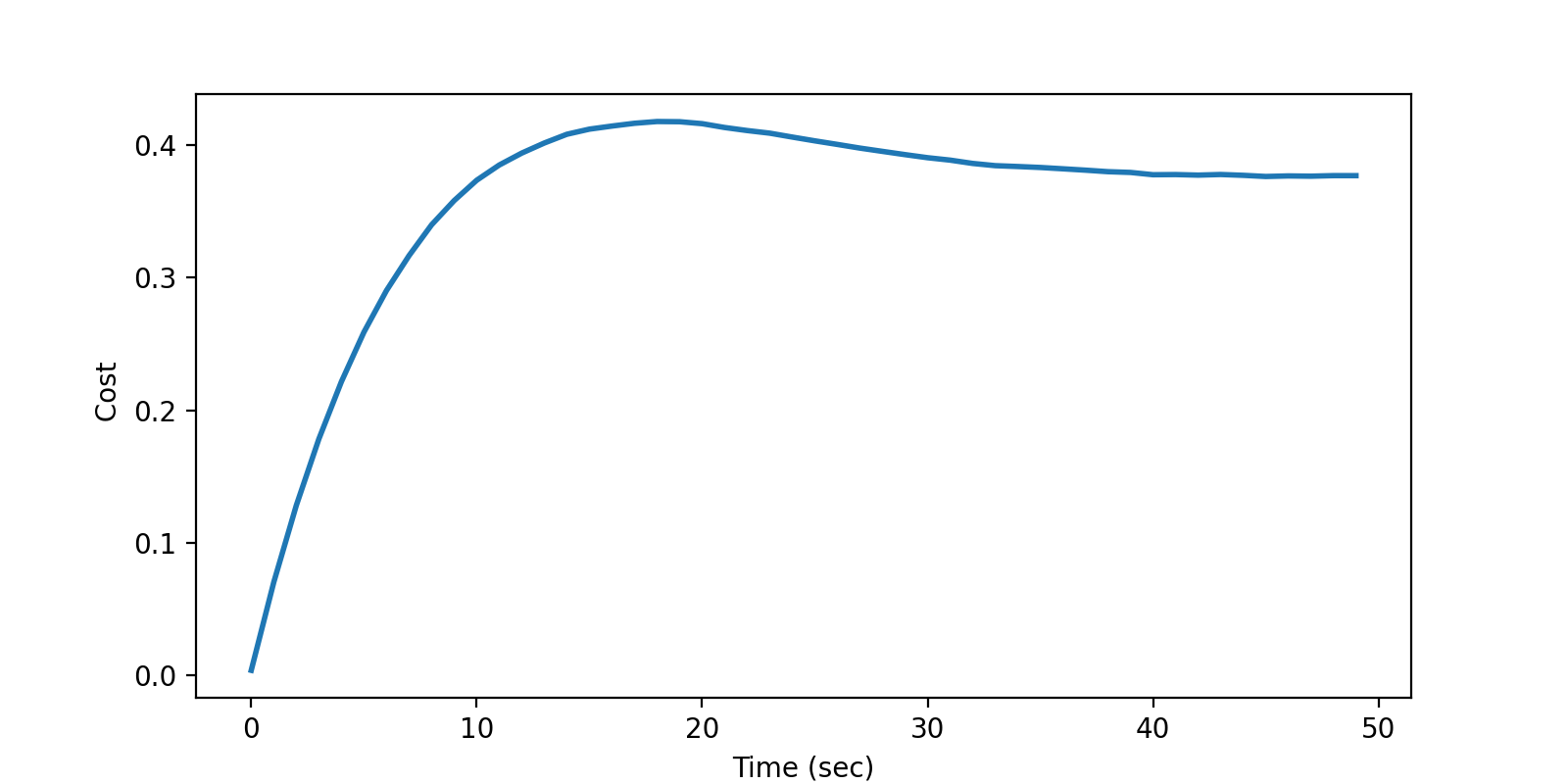}
    	\end{minipage}
	\label{fig:token}
    }    
	\caption{Token experiment.we use blue line for \emph{eCo2}, red line for \emph{TVOC}, yellow line for \emph{AvgCo2} and green line for \emph{AvgTVOC}}
	\label{fig: Trend}	
\end{figure*}

These simulations show, qualitatively, that high levels of compliance with social distancing norms lead to lower infection rates overall. This is the rationale behind the employment of the control laws described in Section II. In, fact as depicted in Figure \ref{fig: Cost_Compliance}, with PFCA including both a global and an individual cost signal, the desired level of compliance is achieved and the fairness of cost among all individuals is also ensured (as each individual will comply equally, regarding of their initial proclivity $q_i$).\newline

\emph{Remark:} We want to emphasize that, the agent model used in this paper does not intend to be a realistic simulation of the dynamics of how Covid-19 spreads across a given population. Rather it is intended as a toy model, to showcase how the control algorithm and the smart mask would work in a similar scenario.

\subsection{Hardware-in-the-loop Simulations}

In the previous section, we showed that with the proposed control strategy, the desired compliance is achieved and the probability of people getting infected is reduced. In order to show the way the mask would implement this control strategy we consider a hardware-in-the-loop simulation based on indoor positioning. As shown in Figure \ref{fig: simula_struc}, the hardware-in-the-loop simulation is used for monitoring smart mask wearing position and status for real agent. The position and the status of the mask are
sent both to the IOTA Tangle and the python simulation in which a number of virtual agents are simulated. Furthermore, the mask status from the simulated agents are appended on the IOTA Tangle, from which are read by the FPCA to generate the cost signals.\newline

To detect the smart mask position we make use of indoor positioning. Although there are multiple viable methods, such as  Bluetooth, WiFi (Wireless local area network ) \cite{woo2011application}, BLE (Bluetooth low energy)  \cite{dinh2020smartphone}, RFID (Radio Frequency Identification) \cite{bernardini2020robot}, either their accuracy is too low or the computational complexity is too high for an IoT application. To achieve better accuracy, our method of choice is Ultra-Wide Band (UWB) due to its better performances for indoor localization, (i.e., high precision and reliable ranging with up to 10 cm accuracy), and large bandwidth with 1 kHz refresh rate \cite{zhu2020adapted}.\newline

\begin{figure}[H]
\centering
\includegraphics[width=0.8\columnwidth]{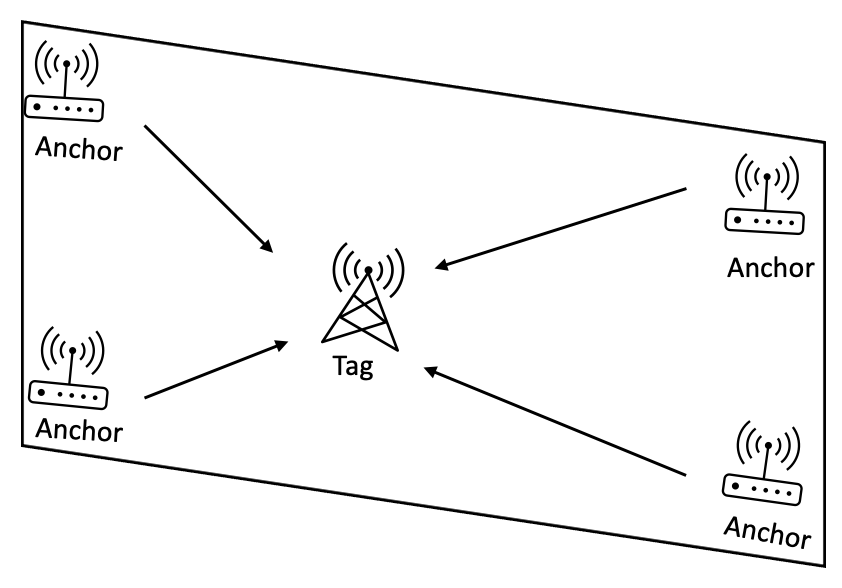}
\caption{Indoor positioning system mode}
\label{fig: Positioning}
\end{figure}

The equipment used in this experiment to perform indoor positioning is the DW1001-DEV\footnote{https://www.decawave.com}. 
We introduce, the concepts of 'Anchor' and 'Tag' to denote different UWB nodes in the system. 'Anchor' represents a fixed node whose position is known. Anchors use wireless signals from tags to determine the position of movable tags whose position is unknown. In this experiment, the  architecture for indoor positioning is depicted in Figure \ref{fig: Positioning}. We set four DWM1001 modules as Anchors and one DWM1001 module as tag (representing the customer moving withing a closed space). Embedded firmware which provides two-way ranging (TWR) and real time location system (RTLS) functionality are pre-loaded in DWM1001 module  \cite{timmurphy.org}. \newline

We adopt TWR method which is based on time of flight (ToF) \cite{sang2018analytical, lian2019numerical} to get the ranging measurements. 
The distance measurement behind this method is that it takes the product of a measured time and the speed of light. The measured time is the radio signal travels between an emitter and a receiver. The principle behind this method is depicted in Figure \ref{fig: Positioning}.
To measure distance, three messages named {\em Poll, Response} and {\em Final} need to be exchanged between anchors and tag. Anchor timestamps including $T_{SP}, T_{SR}, T_{RF}$ and tag timestamps including $T_{SP}, T_{RP}, T_{SF}$ are recorded to calculate the distance. Based on these timestamps, the distance named {\em Dis} between the tag and anchor can be computed by the following equation:
\begin{equation}
\begin{split}
ToF =&\frac{1}{4}*\{(T_{RR}-T_{SP})-(T_{SR}-T_{RP})\\
&+(T_{RF}-T_{SR})-(T_{SF}-T_{RR})\}\\
\end{split}
\end{equation}

\begin{eqnarray}
Dis &= ToF \times c
\end{eqnarray}

where $c$ is the speed of light expressed in $m/s$.


 The experiment is performed in a 20 by 10 m room at Imperial College London. There are four anchors named as DW4105, DW9B10, DW4A2F (as initiator), DW4C15 and one Tag named as DW181C.
In Figure \ref{fig:Realsim1} and \ref{fig:Realsim2}, the triangle depicts the position of the Tag (i.e., the PoC-based mask), while other dots to denote other agents generated by simulator. Figure \ref{fig:co2tvoc} depicts the data collected by the gas sensor.
We average data samples over a fixed time window to detect, whether or not eCo2 and TVOC are above a certain threshold or not. 
More specifically, we set the time window, to be ten samples and we set the thresholds, for the two quantities, respectively, to 500 ppm for eCo2 and 50 ppb for TVOC. According to the designed control algorithm, if the people's mask wearing data can meet for this requirement, which means for agent $i$ $M_i(k)=1$ all staked tokens will be returned. Otherwise when $M_i(k)=0$, all tokens will be deducted. \newline
 
  \textbf{Remark:} This experiment that we have described is designed to illustrate the main features of the operation of the system. In particular, the eCO2 and TVOC thresholds are chosen empirically. Any practical implementation of the system would require these factors to selected in a more rigorous manner by considering factors such as the mask wearers age, health, and factors such as local weather. 
  

During the middle of the time scale when people wear a mask on, the value for all variables are increasing dramatically compared to the rest of value when people do not wear a mask. Figure \ref{fig:token} depicts the changing of individual cost according to the compliance of the social contract -- when wearing masks correctly. 

\section{Conclusion}
In this paper, a smart mask prototype is designed to monitor people's mask wearing status. The use of DLT -- IOTA Tangle, severs as both a communication layer for the control algorithm as well as ledger ensuring the security and immutability of data. The designed mechanism is validated through  extensive  simulations including a python-based one and a hardware-in-the-loop one.

\section*{Acknowledgement}
Thanks for the support from DecaWave company to supply DecaWave DWM1001-DEV.

\bibliographystyle{unsrt}
\bibliography{main}

\end{document}